\documentclass[12pt,draft]{article}
\usepackage{amsfonts,graphics}
\newcommand{\sfrac}[2]{{\textstyle{\frac{#1}{#2}}}}
\newcommand{\half}{{\textstyle{\frac{1}{2}}}}
\newcommand{\thalf}{{\textstyle{\frac{3}{2}}}}
\newcommand{\ZZ}{{\mathbb Z}}

\newcommand{\llra}{\leftrightarrow}

\newtheorem{thm}{Theorem}

\begin{document}
\newpage
\pagestyle{empty}
\setcounter{page}{0}
\vfill
\begin{center}

{\Large \textbf{Crystalizing the Genetic Code}}

\vspace{10mm}

{\large L. Frappat$^a$, A. Sciarrino$^{b,a}$, P. Sorba$^a$}

\vspace{10mm}

\emph{$^a$ Laboratoire d'Annecy-le-Vieux de Physique Th{\'e}orique LAPTH}

\emph{CNRS, UMR 5108, associ{\'e}e {\`a} l'Universit{\'e} de Savoie}

\emph{BP 110, F-74941 Annecy-le-Vieux Cedex, France}

\vspace{7mm}

\emph{$^b$ Permanent adress: Dipartimento di Scienze Fisiche,
Universit{\`a} di Napoli ``Federico II''}

\emph{and I.N.F.N., Sezione di Napoli}

\emph{Complesso di Monte S. Angelo, Via Cintia, I-80126 Napoli, Italy} 

\end{center}

\vspace{12mm}

\begin{abstract}
New developments are presented in the framework of the model introduced by 
the authors in refs.  \cite{FSS1,FSS2} and in which nucleotides as well as 
codons are classified in crystal bases of the quantum group $U_{q}(sl(2) 
\oplus sl(2))$ in the limit $q \to 0$.  An operator which gives the 
correspondence between the amino-acids and the codons is now obtained for 
any known genetic code.  The free energy released by base pairing of 
dinucleotides as well as the relative hydrophilicity and hydrophobicity of 
the dinucleosides are also computed.  For the vertebrate series, a 
universal behaviour in the ratios of codon usage frequencies is put in 
evidence and is shown to fit nicely in our model.  Then a first attempt to 
represent the mutations relative to the deletion of a pyrimidine by action 
of a suitable crystal spinor operator is proposed.  Finally recent 
theoretical descriptions are reviewed and compared with our model.
\end{abstract}

\vfill
\vfill

\vfill
\vfill

\rightline{LAPTH-787/00}
\rightline{DSF-TH-9/00}
\rightline{physics/0003037}
\rightline{March 2000}

\clearpage
\pagestyle{plain}
\baselineskip=18pt

\section{Introduction}

Among the numerous and important questions offered to the physicist by the 
sciences of life, the ones relative to the genetic code present a 
particular interest.  Indeed, in addition to the fundamental importance of 
this domain, the DNA structure on the one hand and the mechanism of 
polypeptid fixation from codons on the other hand possess appealing aspects 
for the theorist.  Let us, in a brief summary, select some essential 
features \cite{livre}.  First, as well known, the DNA macromolecule is 
constituted by two linear chains of nucleotides in a double helix shape.  
There are four different nucleotides, characterized by their bases: adenine 
(A) and guanine (G) deriving from purine, and cytosine (C) and thymine (T) 
coming from pyrimidine.  Note also that an A (resp.  T) base in one strand 
is connected with two hydrogen bonds to a T (resp.  A) base in the other 
strand, while a C (resp.  G) base is related to a G (resp.  C) base with 
three hydrogen bonds.  The genetic information is transmitted to the 
cytoplasm via the messenger ribonucleic acid or mRNA. During this 
operation, called transcription, the A, G, C, T bases in the DNA are 
associated respectively to the U, C, G, A bases, U denoting the uracile 
base.  Then it will be through a ribosome that a triplet of nucleotides or 
codon will be related to an amino-acid.  More precisely, a codon is defined 
as an ordered sequence of three nucleotides, e.g. AAG, ACG, etc., and one 
enumerates in this way $4 \!  \times \!  4 \!  \times \!  4 = 64$ different 
codons.  Following the universal eukariotic code (see Table 
\ref{tablerep}), 61 of such triplets can be connected in an unambiguous way 
to the amino-acids, except the three following triplets UAA, UAG and UGA, 
which are called non-sense or stop-codons, the role of which is to stop the 
biosynthesis.  Indeed, the genetic code is the association between codons 
and amino-acids.  But since one distinguishes only 20 
amino-acids\footnote{Alanine (Ala), Arginine (Arg), Asparagine (Asn), 
Aspartic acid (Asp), Cysteine (Cys), Glutamine (Gln), Glutamic acid (Glu), 
Glycine (Gly), Histidine (His), Isoleucine (Ile), Leucine (Leu), Lysine 
(Lys), Methionine (Met), Phenylalanine (Phe), Proline (Pro), Serine (Ser), 
Threonine (Thr), Tryptophane (Trp), Tyrosine (Tyr), Valine (Val).} related 
to the 61 codons, it follows that the genetic code is degenerated.  Still 
considering the standard eukariotic code, one observes sextets, 
quadruplets, triplets, doublets and singlets of codons, each multiplet 
corresponding to a specific amino-acid.  Such a picture naturally suggests 
to look for an underlying symmetry able to describe the observed structure 
in multiplets, in the spirit of dynamical symmetry scheme which has proven 
so powerful in atomic, molecular and nuclear physics.  We review at the end 
of this paper these recent approaches.

\medskip

In refs.  \cite{FSS1,FSS2} we have proposed a mathematical framework in 
which the codons appear as composite states of nucleotides.  The four 
nucleotides being assigned to the fundamental irreducible representation of 
the quantum group ${\cal U}_{q}(sl(2) \oplus sl(2))$ in the limit $q \to 
0$, the codons are obtained as tensor product of nucleotides.  Indeed, the 
properties of quantum group representations in the limit $q \to 0$, or 
crystal basis, are well adapted to take into account the nucleotide 
ordering.  Then properties of this model have been considered.  We will 
generalize some of them in the following and also propose new developments.

\medskip

The paper is organized as follows.  We start in sect.  \ref{model} by 
recalling the main aspects of the model.  In sect.  \ref{reading} we build 
out of the generators of ${\cal U}_{q \to 0}(sl(2) \oplus sl(2))$ a 
reading operator, which gives the correct correspondence between codons 
and amino-acids for each of the 12 presently known genetic codes.  This 
construction generalizes in a synthetical way the one started in 
\cite{FSS1} for the eukariotic and vertebrate mitochondrial codes, the 
different reading operators acting on codons and providing the same 
eigenvalue for a given amino-acid whatever the considered code.  In sect.  
\ref{fits} some physical properties of dinucleotide states are fitted.  In 
sect.  \ref{correl}, we analyze ratios of codon usage frequency for several 
biological species belonging to the vertebrate class and put in evidence a 
universal behaviour, which fits naturally in our model.  In sect.  
\ref{mutation}, making use of the general crystal basis mathematical 
framework, we represent the mutation induced by the deletion of a 
pyrimidine by the action of a suitable crystal spinor operator.  In sect.  
\ref{compare} we review and compare with our model the recent symmetry 
approaches to the genetic code.  Finally in sect.  \ref{conclusion} we give 
a few conclusions and discuss some directions of future developments.
 
\section{The Model}
\label{model}

We consider the four nucleotides as basic states of the $(\half,\half)$ 
representation of the ${\cal U}_{q}(sl(2) \oplus sl(2))$ quantum enveloping 
algebra in the limit $q \to 0$.  A triplet of nucleotides will then be 
obtained by constructing the tensor product of three such four-dimensional 
representations.  Actually, this approach mimicks the group theoretical 
classification of baryons made out from three quarks in elementary 
particles physics, the building blocks being here the A, C, G, T/U 
nucleotides.  The main and essential difference stands in the property of a 
codon to be an \emph{ordered} set of three nucleotides, which is not the 
case for a baryon.

\medskip

Constructing such pure states is made possible in the framework of any 
algebra ${\cal U}_{q \to 0}({\cal G})$ with ${\cal G}$ being any 
(semi)-simple classical Lie algebra owing to the existence of a special 
basis, called crystal basis, in any (finite dimensional) representation of 
${\cal G}$.  The algebra ${\cal G} = sl(2) \oplus sl(2)$ appears the most 
natural for our purpose.  The complementary rule in the DNA--mRNA 
transcription may suggest to assign a \emph{quantum number} with opposite 
values to the couples (A,T/U) and (C,G).  The distinction between the 
purine bases (A,G) and the pyrimidine ones (C,T/U) can be algebraically 
represented in an analogous way.  Thus considering the fundamental 
representation $(\half,\half)$ of $sl(2) \oplus sl(2)$ and denoting $\pm$ 
the basis vector corresponding to the eigenvalues $\pm\half$ of the $J_{3}$ 
generator in any of the two $sl(2)$ corresponding algebras, we will assume 
the following ``biological'' spin structure:
\begin{eqnarray}
	&sl(2)_{H}& \nonumber \\
	C \equiv (+,+) &\qquad\longleftrightarrow\qquad& U \equiv (-,+) 
	\nonumber \\
	\Bigg.  sl(2)_{V} \updownarrow && \updownarrow sl(2)_{V} \\
	\Bigg.  G \equiv (+,-) &\qquad\longleftrightarrow\qquad& A \equiv (-,-) 
	\nonumber \\
	&sl(2)_{H}& \nonumber \\
	\label{eq:gc1}
\end{eqnarray}
the subscripts $H$ (:= horizontal) and $V$ (:= vertical) being just 
added to specify the algebra.

\medskip

Now, we consider the representations of ${\cal U}_{q}(sl(2))$ and 
more specifically the crystal bases obtained when $q \to 0$.  Introducing 
in ${\cal U}_{q \to 0}(sl(2))$ the operators $J_{+}$ and $J_{-}$ after 
modification of the corresponding simple root vectors of ${\cal 
U}_{q}(sl(2))$, a particular kind of basis in a ${\cal 
U}_{q}(sl(2))$-module can be defined.  Such a basis is called a crystal 
basis and carries the property to undergo in a specially simple way the 
action of the $J_{+}$ and $J_{-}$ operators: as an example, for any couple 
of vectors $u,v$ in the crystal basis ${\cal B}$, one gets $u = J_{+} v$ if 
and only if $v = J_{-} u$.  More interesting for our purpose is the crystal 
basis in the tensorial product of two representations.  Then the following 
theorem holds \cite{Kashi} (written here in the case of $sl(2)$):
\begin{thm}[Kashiwara]
	Let ${\cal B}_{1}$ and ${\cal B}_{2}$ be the crystal bases of the 
	$M_{1}$ and $M_{2}$ ${\cal U}_{q \to 0} (sl(2))$-modules 
	respectively.  Then for $u \in {\cal B}_{1}$ and $v \in {\cal B}_{2}$, 
	we have:
	\begin{eqnarray}
		&& J_{-}(u \otimes v) = \left\{
		\begin{array}{ll}
			J_{-} u \otimes v & \exists \, n \ge 1 \mbox{ such that } 
			J_{-}^n u \ne 0 \mbox{ and } J_{+} v = 0 \\
			u \otimes J_{-} v & \mbox{otherwise} \\
		\end{array} \right.  \\
		&& J_{+} (u \otimes v) = \left\{
		\begin{array}{ll}
			u \otimes J_{+} v & \exists \, n \ge 1 \mbox{ such that } 
			J_{+}^n v \ne 0 \mbox{ and } J_{-} u = 0 \\
			J_{+} u \otimes v & \mbox{otherwise} \\
		\end{array} \right.
	\end{eqnarray}
\end{thm}

Note that the tensor product of two representations in the crystal basis is 
not commutative.  However, in the case of our model, we only need to 
construct the $n$-fold tensor product of the fundamental representation 
$(\half,\half)$ of ${\cal U}_{q \to 0}(sl(2) \oplus sl(2))$ by itself, thus 
preserving commutativity and associativity.

Let us insist on the choice of the crystal basis, which exists only in the 
limit $q \to 0$.  In a codon the order of the nucleotides is of fundamental 
importance (e.g. CCU $\to$ \texttt{Pro}, CUC $\to$ \texttt{Leu}, UCC $\to$ 
\texttt{Ser}).  If we want to consider the codons as composite states of 
the (elementary) nucleotides, this surely cannot be done in the framework 
of Lie (super)algebras.  Indeed in the Lie theory, the composite states are 
obtained by performing tensor products of the fundamental irreducible 
representations.  They appear as linear combinations of the elementary 
states, with symmetry properties determined from the tensor product (i.e. 
for $sl(n)$, by the structure of the corresponding Young tableaux).  On the 
contrary the crystal basis provides us with the mathematical structure to 
build composite states as \emph{pure} states, characterized by the order of 
the constituents.  In order to dispose of such a basis, we need to consider 
the limit $q \to 0$.  Note that in this limit we do not deal anymore either 
with a Lie algebra or with an universal deformed enveloping algebra.

To represent a codon, we have to perform the tensor product of three 
$(\half,\half)$ representations of ${\cal U}_{q \to 0}(sl(2) \oplus 
sl(2))$.  However, it is well-known (see Tables \ref{tablerep}) that in a 
multiplet of codons relative to a specific amino-acid, the two first bases 
constituent of a codon are ``relatively stable'', the degeneracy being 
mainly generated by the third nucleotide.  We consider first the tensor 
product:
\begin{equation}
	(\half,\half) \, \otimes \, (\half,\half) = (1,1) \, \oplus \, (1,0) \, 
	\oplus \, (0,1) \, \oplus \, (0,0)
	\label{eq:gc4}
\end{equation}
where inside the parenthesis, $j=0,\half,1$ is put in place of the 
$2j+1=1,2,3$ respectively dimensional $sl(2)$ representation.  We get, 
using Theorem 1, the following tableau:
\begin{displaymath}
	\begin{array}{lcccc}
		\to \,\, su(2)_H \qquad \qquad &(0,0) & (\mbox{CA}) &\qquad 
		\qquad (1,0) & (
		\begin{array}{ccc} 
			\mbox{CG} & \mbox{UG} & \mbox{UA} \\
		\end{array}) \\
		\downarrow \\
		su(2)_V &(0,1) & \left(
		\begin{array}{c}
			\mbox{CU} \\
			\mbox{GU} \\
			\mbox{GA} \\
		\end{array}
		\right) & \qquad \qquad (1,1) & \left(
		\begin{array}{ccc}
			\mbox{CC} & \mbox{UC} & \mbox{UU} \\
			\mbox{GC} & \mbox{AC} & \mbox{AU} \\
			\mbox{GG} & \mbox{AG} & \mbox{AA} \\
		\end{array}
		\right)
	\end{array}
\end{displaymath}
{From} Table \ref{tablerep}, the dinucleotide states formed by the first 
two nucleotides in a codon can be put in correspondence with quadruplets, 
doublets or singlets of codons relative to an amino-acid.  Note that the 
sextets (resp.  triplets) are viewed as the sum of a quadruplet and a 
doublet (resp.  a doublet and a singlet).  Let us define the ``charge'' $Q$ 
of a dinucleotide state by
\begin{equation}
	Q = J_{H,3}^{(1)} + J_{H,3}^{(2)} + J_{V,3}^{(2)}
	\label{eq:charge}
\end{equation}
where the superscript $(1)$ or $(2)$ denotes the position of a codon in the 
dinucleotide state.  \\
The dinucleotide states are then split into two octets with respect to the 
charge $Q$: the eight \emph{strong} dinucleotides associated to the 
quadruplets (as well as those included in the sextets) of codons satisfy $Q 
> 0$, while the eight \emph{weak} dinucleotides associated to the doublets 
(as well as those included in the triplets) and eventually to the singlets 
of codons satisfy $Q < 0$.  Let us remark that by the change $ C \llra A $ 
and $ U \llra G $, which is equivalent to the change of the sign of $J_{3, 
\alpha}$ or to reflexion with respect to the diagonals of the 
eq.(\ref{eq:gc1}), the 8 strong dinucleotides are transformed into weak 
ones and vice-versa.

\medskip

If we consider the three-fold tensor product, the content into irreducible 
representations of ${\cal U}_{q \to 0}(sl(2) \oplus sl(2))$ is given by:
\begin{equation}
	(\half,\half) \otimes (\half,\half) \otimes (\half,\half) = 
	(\thalf,\thalf) \oplus 2 \, (\thalf,\half) \oplus 2 \, (\half,\thalf) 
	\oplus 4 \, (\half,\half)
	\label{eq:gc5}
\end{equation}
The structure of the irreducible representations of the r.h.s.  of Eq.  
(\ref{eq:gc5}) is (the upper labels denote different irreducible 
representations):
\begin{eqnarray*}
	&(\thalf,\thalf) \equiv \left( 
	\begin{array}{cccc} 
		\mbox{CCC} & \mbox{UCC} & \mbox{UUC} & \mbox{UUU} \\
		\mbox{GCC} & \mbox{ACC} & \mbox{AUC} & \mbox{AUU} \\
		\mbox{GGC} & \mbox{AGC} & \mbox{AAC} & \mbox{AAU} \\
		\mbox{GGG} & \mbox{AGG} & \mbox{AAG} & \mbox{AAA} \\
	\end{array}
	\right)& \\
	&& \\
	&(\thalf,\half)^{1} \equiv \left( 
	\begin{array}{cccc} 
		\mbox{CCG} & \mbox{UCG} & \mbox{UUG} & \mbox{UUA} \\
		\mbox{GCG} & \mbox{ACG} & \mbox{AUG} & \mbox{AUA} \\
	\end{array}
	\right)& \\
	&& \\
	&(\thalf,\half)^{2} \equiv \left( 
	\begin{array}{cccc} 
		\mbox{CGC} & \mbox{UGC} & \mbox{UAC} & \mbox{UAU} \\
		\mbox{CGG} & \mbox{UGG} & \mbox{UAG} & \mbox{UAA} \\
	\end{array}
	\right)& \\
	&& \\
	&(\half,\thalf)^{1} \equiv \left( 
	\begin{array}{cc}
		\mbox{CCU} &\mbox{UCU} \\
		\mbox{GCU} & \mbox{ACU} \\
		\mbox{GGU} & \mbox{AGU} \\
		\mbox{GGA} & \mbox{AGA} \\
	\end{array} 
	\right)
	\qquad \qquad
	(\half,\thalf)^{2} \equiv \left( 
	\begin{array}{cc}
		\mbox{CUC} & \mbox{CUU} \\
		\mbox{GUC} & \mbox{GUU} \\
		\mbox{GAC} & \mbox{GAU} \\ 
		\mbox{GAG} & \mbox{GAA} \\
	\end{array} 
	\right)& \\
	&& \\
	&(\half,\half)^{1} \equiv \left( 
	\begin{array}{cc}
		\mbox{CCA} & \mbox{UCA} \\ 
		\mbox{GCA} & \mbox{ACA} \\ 
	\end{array} 
	\right)
	\qquad \qquad
	(\half,\half)^{2} \equiv \left( 
	\begin{array}{cc}
		\mbox{CGU} & \mbox{UGU} \\
		\mbox{CGA} & \mbox{UGA} \\
	\end{array}
	\right)& \\
	&& \\
	&(\half,\half)^{3} \equiv \left( 
	\begin{array}{cc}
		\mbox{CUG} & \mbox{CUA} \\ 
		\mbox{GUG} & \mbox{GUA} \\ 
	\end{array} 
	\right)
	\qquad \qquad
	(\half,\half)^{4} \equiv \left( 
	\begin{array}{cc}
		\mbox{CAC} & \mbox{CAU} \\ 
		\mbox{CAG} & \mbox{CAA} \\ 
	\end{array} 
	\right)&
\end{eqnarray*}

\medskip

The correspondence with the amino-acids is given in Table \ref{euka} (for 
the eukariotic code).  

\begin{figure}
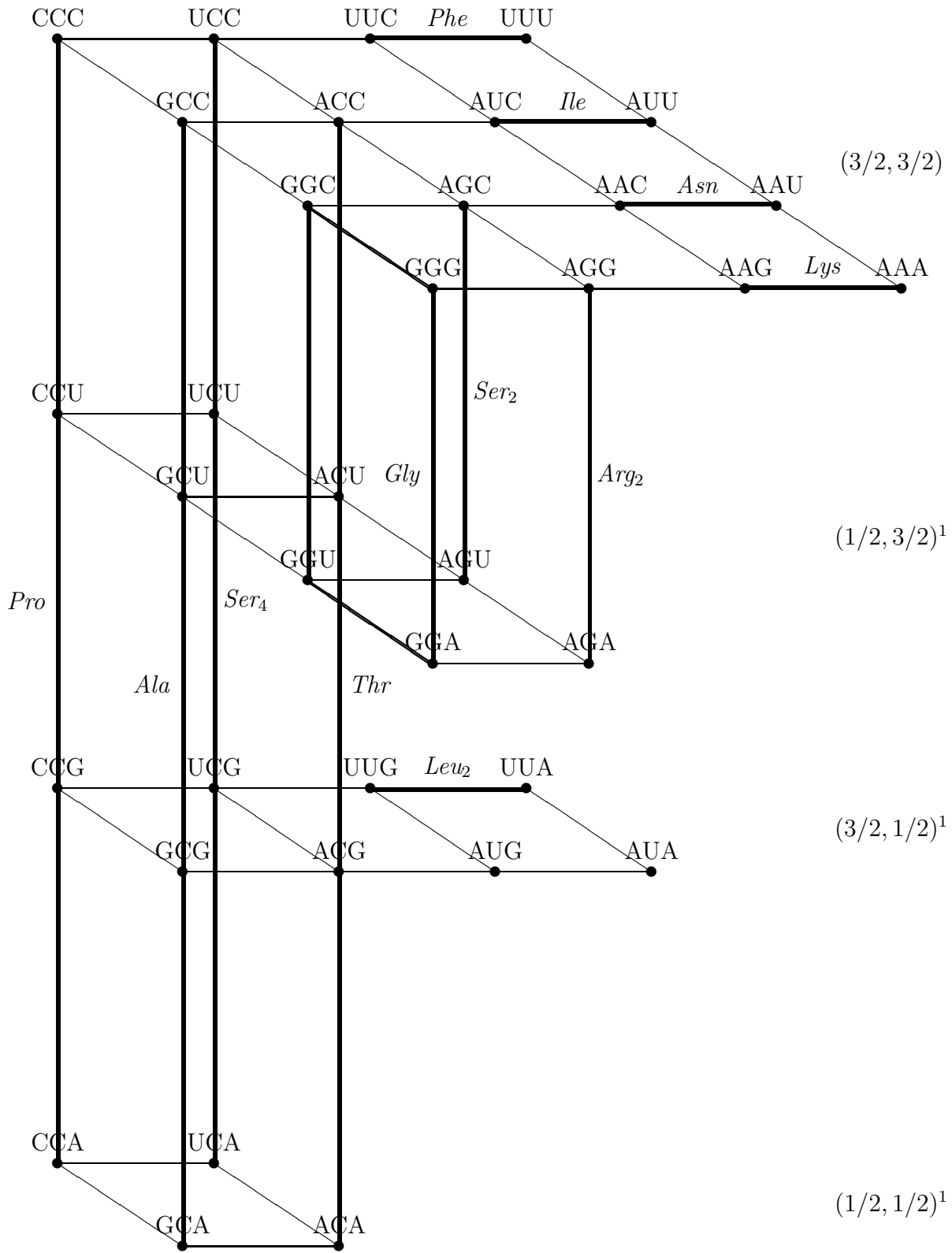

\caption{Classification of the codons in the different crystal bases.}
\label{figA}
\end{figure}

Let us close this section by drawing the reader's attention to Fig.  
\ref{figA} where is specified for each codon its position in the 
appropriate representation.  The diagram of states for each representation 
is supposed to lie in a separate parallel plane.  Thick lines connect 
codons associated to the same amino-acid.  One remarks that each segment 
relates a couple of codons belonging to the same representation or to two 
different representations.  This last case occurs for quadruplets or 
sextets of codons associated to the same amino-acid.

\section{The Reading (or Ribosome) operator ${\cal R}$}
\label{reading}

\subsection{General structure of the reading operator}

As expected from formula (\ref{eq:gc5}), our model does not gather 
codons associated to one particular amino-acid in the same irreducible 
multiplet.  However, it is possible to construct an operator ${\cal R}$ out 
of the algebra ${\cal U}_{q \to 0}(sl(2) \oplus sl(2))$, acting on the 
codons, that will describe the various genetic codes in the following way:

\emph{Two codons have the same eigenvalue under ${\cal R}$ if and only if they 
are associated to the same amino-acid.  This operator ${\cal R}$ will be 
called the reading operator.}

\medskip

It is a remarkable fact that the various genetic codes share the same basic 
structure.  As we mentioned above, the dinucleotides can be split into 
``strong'' dinucleotides CC, GC, UC, AC, CU, GU, CG and GG that lead to 
quartets and ``weak'' ones UU, AU, UG, AG, CA, GA, UA, AA that lead to 
doublets.  Let us construct a prototype of the reading operator that 
reproduces this structure.

\medskip

\noindent The first part of the reading operator ${\cal R}$ is responsible for 
the structure in quadruplets given essentially by the dinucleotide content.  
It is given by (the $c_{i}$ are arbitrary coefficients)
\begin{equation}
	\sfrac{4}{3} c_{1} \, C_{H} + \sfrac{4}{3} c_{2} \, C_{V} - 4c_{1} \, 
	{\cal P}_{H} \, J_{H,3} - 4c_{2} \, {\cal P}_{V} \, J_{V,3} \;.
	\label{eq:readop1}
\end{equation}
The operators $J_{\alpha,3}$ ($\alpha = H,V$) are the third components of 
the total spin generators of the algebra ${\cal U}_{q \to 0}(sl(2) \oplus 
sl(2))$.  The operator $C_{\alpha}$ is a Casimir operator of ${\cal U}_{q 
\to 0}(sl(2)_{\alpha})$ in the crystal basis.  It commutes with 
$J_{\alpha\pm}$ and $J_{\alpha,3}$ and its eigenvalues on any vector basis 
of an irreducible representation of highest weight $J$ is $J(J+1)$, that is 
the same as the undeformed standard second degree Casimir operator of 
$sl(2)$.  Its explicit expression is
\begin{equation}
	C_{\alpha} = (J_{\alpha,3})^{2} + \half \sum_{n \in \ZZ_+} \sum_{k=0}^n 
	(J_{\alpha-})^{n-k} (J_{\alpha+})^n (J_{\alpha-})^k \;.
\end{equation}
Note that for $sl(2)_{q \to 0}$ the Casimir operator is an infinite series 
of powers of $J_{\alpha\pm}$.  However in any finite irreducible 
representation only a finite number of terms gives a non-vanishing 
contribution.  \\
${\cal P}_{H}$ and ${\cal P}_{V}$ are projectors given by the following 
expressions:
\begin{equation}
	{\cal P}_{H} = J_{H+}^d \, J_{H-}^d \qquad \mbox{and} \qquad {\cal 
	P}_{V} = J_{V+}^d \, J_{V-}^d \;.
	\label{eq:proj1}
\end{equation}

\medskip

\noindent The second part of ${\cal R}$ gives rise to the splitting of the 
quadruplets into doublets.  It reads
\begin{equation}
	-2 {\cal P}_{D} \, c_{3} \, J_{V,3}
	\label{eq:readop2}
\end{equation}
where the projector ${\cal P}_{D}$ is given by
\begin{eqnarray}
	{\cal P}_{D} &\!\!=\!\!& (1-J_{V+}^d \, J_{V-}^d) (J_{H+}^d \, 
	J_{H-}^d) (J_{H-}^d \, J_{H+}^d) + (1-J_{H+}^d \, J_{H-}^d) (1-J_{V+}^d 
	\, J_{V-}^d) \nonumber \\
	&& + \, (1-J_{H+}^d \, J_{H-}^d) (J_{V+}^d \, J_{V-}^d) (J_{H-}^d \, 
	J_{H+}^d) \;.
	\label{eq:proj2}
\end{eqnarray}

\medskip

\noindent The third part of ${\cal R}$ allows to reproduce the sextets viewed 
as quartets plus doublets.  It is
\begin{equation}
	-2 {\cal P}_{S} \, c_{4} \, J_{V,3} 
	\label{eq:readop3}
\end{equation}
where the projector ${\cal P}_{S}$ is given by
\begin{equation}
	{\cal P}_{S} = (J_{H-}^d \, J_{H+}^d) \,\, [(J_{H+}^d \, J_{H-}^d) 
	(1-J_{V+}^d \, J_{V-}^d) + (J_{V+}^d \, J_{V-}^d)(J_{V-}^d \, J_{V+}^d) 
	(1-J_{H+}^d \, J_{H-}^d )] \;.
	\label{eq:proj3}
\end{equation}

\medskip

\noindent At this point, one obtains the eigenvalues of the reading 
operator ${\cal R}$ for the 64 codons, where Y = C,U (pyrimidines), R = G,A 
(purines) and N = C,U,G,A:
\begin{equation}
	\begin{array}{lcl}
		\mbox{CCN} = - c_{1} - c_{2} &\quad&
		\mbox{GCN} = - c_{1} + 3 c_{2} \\
		\mbox{UCN} = 3 c_{1} - c_{2} &\quad&
		\mbox{ACN} = 3 c_{1} + 3 c_{2} \\
		\mbox{CUN} = c_{1} - c_{2} &\quad&
		\mbox{GUN} = c_{1} + 3 c_{2} \\
		\mbox{CGN} = - c_{1} + c_{2} &\quad&
		\mbox{GGN} = - c_{1} + 5 c_{2} \\
		\mbox{UUY} = 5 c_{1} - c_{2} - 3 c_{3} &\quad&
		\mbox{UUR} = 5 c_{1} - c_{2} - c_{3} \\
		\mbox{AUY} = 5 c_{1} + 3 c_{2} - c_{3} - c_{4} &\quad&
		\mbox{AUR} = 5 c_{1} + 3 c_{2} + c_{3} + c_{4} \\
		\mbox{UGY} = 3 c_{1} + c_{2} - c_{3} - c_{4} &\quad&
		\mbox{UGR} = 3 c_{1} + c_{2} + c_{3} + c_{4} \\
		\mbox{AGY} = 3 c_{1} + 5 c_{2} + c_{3} + c_{4} &\quad&
		\mbox{AGR} = 3 c_{1} + 5 c_{2} + 3 c_{3} + 3 c_{4} \\
		\mbox{CAY} = c_{1} + c_{2} - c_{3} &\quad&
		\mbox{CAR} = c_{1} + c_{2} + c_{3} \\
		\mbox{GAY} = c_{1} + 5 c_{2} + c_{3} &\quad&
		\mbox{GAR} = c_{1} + 5 c_{2} + 3 c_{3} \\
		\mbox{UAY} = 5 c_{1} + c_{2} - c_{3} &\quad&
		\mbox{UAR} = 5 c_{1} + c_{2} + c_{3} \\
		\mbox{AAY} = 5 c_{1} + 5 c_{2} + c_{3} &\quad&
		\mbox{AAR} = 5 c_{1} + 5 c_{2} + 3 c_{3} \\
	\end{array}
\end{equation}
The coefficients $c_{3}$ and $c_{4}$ are fixed as follows.  The coefficient 
$c_{3}$ is set to the value $c_{3} = 4c_{1}$ by requiring that the quartet 
CUN and the doublet UUR, associated to the amino-acid \texttt{Leu}, lead to 
the same ${\cal R}$-eigenvalue.  It remains to reproduce the \texttt{Ser} 
sextet.  This is achieved by taking for the coefficient $c_{4}$ the value 
$c_{4} = -4c_{1}-6c_{2}$, such that the final eigenvalues for the codons 
are the following:
\begin{equation}
	\begin{array}{lclclcl}
		\mbox{CCN} = - c_{1} - c_{2} &\quad&
		\mbox{GCN} = - c_{1} + 3 c_{2} &\quad&
		\mbox{UCN} = 3 c_{1} - c_{2} &\quad&
		\mbox{ACN} = 3 c_{1} + 3 c_{2} \\
		\mbox{CUN} = c_{1} - c_{2} &\quad&
		\mbox{GUN} = c_{1} + 3 c_{2} &\quad& 
		\mbox{CGN} = c_{1} + c_{2} &\quad&
		\mbox{GGN} = - c_{1} + 5 c_{2} \\
		\mbox{UUY} = - 7 c_{1} - c_{2} &\quad&
		\mbox{UUR} = c_{1} - c_{2} &\quad& 
		\mbox{AUY} = 5 c_{1} + 9 c_{2} &\quad&
		\mbox{AUR} = 5 c_{1} - 3 c_{2} \\
		\mbox{UGY} = 3 c_{1} + 7 c_{2} &\quad&
		\mbox{UGR} = 3 c_{1} - 5 c_{2} &\quad& 
		\mbox{AGY} = 3 c_{1} - c_{2} &\quad&
		\mbox{AGR} = 3 c_{1} - 13 c_{2} \\
		\mbox{CAY} = - 3 c_{1} + c_{2} &\quad&
		\mbox{CAR} = 5 c_{1} + c_{2} &\quad& 
		\mbox{GAY} = 5 c_{1} + 5 c_{2} &\quad&
		\mbox{GAR} = 13 c_{1} + 5 c_{2} \\
		\mbox{UAY} = c_{1} + c_{2} &\quad&
		\mbox{UAR} = 9 c_{1} + c_{2} &\quad& 
		\mbox{AAY} = 9 c_{1} + 5 c_{2} &\quad&
		\mbox{AAR} = 17 c_{1} + 5 c_{2} \\
	\end{array}
\end{equation}
The prototype of the reading operator ${\cal R}$ takes finally the form:
\begin{eqnarray}
	{\cal R} = \sfrac{4}{3} c_{1} \, C_{H} + \sfrac{4}{3} c_{2} \, C_{V} - 
	4c_{1} \, {\cal P}_{H} \, J_{H,3} - 4c_{2} \, {\cal P}_{V} \, J_{V,3} + 
	(-8c_{1} \, {\cal P}_{D} \, + (8c_{1}+12c_{2}) \, {\cal P}_{S}) \, 
	J_{V,3}
	\label{eq:readopproto}
\end{eqnarray}
and the correspondence codons/amino-acids is given as follows:
\begin{equation}
	\begin{tabular}{lllllll}
		CCN $\to$ \texttt{Pro} & \qquad & UCN $\to$ \texttt{Ser} & \qquad & 
		GCN $\to$ \texttt{Ala} & \qquad & ACN $\to$ \texttt{Thr} \\
		CUN $\to$ \texttt{Leu} & \qquad & GUN $\to$ \texttt{Val} & \qquad & 
		CGN $\to$ \texttt{Arg} & \qquad & GGN $\to$ \texttt{Gly} \\
		UUY $\to$ \texttt{Phe} & \qquad & AUY $\to$ \texttt{Ile} & \qquad & 
		UGY $\to$ \texttt{Cys} & \qquad & AGY $\to$ \texttt{Ser} \\
		UUR $\to$ \texttt{Leu} & \qquad & AUR $\to$ \texttt{Met} & \qquad & 
		UGR $\to$ \texttt{Trp} & \qquad & AGR $\to$ unassigned (X) \\
		CAY $\to$ \texttt{His} & \qquad & UAY $\to$ \texttt{Tyr} & \qquad & 
		GAY $\to$ \texttt{Gln} & \qquad & AAY $\to$ \texttt{Asn} \\
		CAR $\to$ \texttt{Gln} & \qquad & UAR $\to$ \texttt{Ter} & \qquad & 
		GAR $\to$ \texttt{Glu} & \qquad & AAR $\to$ \texttt{Lys} \\
	\end{tabular}
	\label{eq:codes}
\end{equation}

\subsection{The various genetic codes}

In this section, we will determine the reading operators for the following 
genetic codes: \\
-- the Eukariotic Code (EC), \\
-- the Vertebral Mitochondrial Code (VMC), \\
-- the Yeast Mitochondrial Code (YMC), \\
-- the Invertebrate Mitochondrial Code (IMC), \\
-- the Protozoan Mitochondrial and Mycoplasma Code (PMC), \\
-- the Echinoderm Mitochondrial Code (EMC), \\
-- the Ascidian Mitochondrial Code (AMC), \\
-- the Flatworm Mitochondrial Code (FMC), \\
-- the Ciliate Nuclear Code (CNC), \\
-- the Blepharisma Nuclear Code (BNC), \\
-- the Euplotid Nuclear Code (ENC), \\
-- the Alternative Yeast Nuclear Code (alt.  YNC), \\
Let us emphasize that each of these codes is very close to the assignment 
(\ref{eq:codes}). The main differences between the biological codes and the 
prototype code (\ref{eq:codes}) are the following:
\begin{itemize}
	\item
	assignment of the doublet AGR either to \texttt{Arg} (codes EC, YMC, 
	PMC, CNC, BNC, ENC, aYNC), \texttt{Ser} (codes IMC, EMC, FMC), 
	\texttt{Gly} (code AMC) or the stop signal \texttt{Ter} (code VMC).  \\
	Such an assignment is done by the following term in the reading 
	operator:
	\begin{equation}
		c_{5} \; {\cal P}_{AG} \; \Big( \half - J_{V,3}^{(3)} \Big)
		\label{eq:readop5}
	\end{equation}
	The operators $J_{\alpha,3}^{(3)}$ are the third components 
	corresponding to the third nucleotide of a codon.  Of course, these 
	last two operators can be replaced by $J_{\alpha,3}^{(3)} = 
	J_{\alpha,3} - J_{\alpha,3}^d$.  \\
	The projector ${\cal P}_{AG}$ is given by
	\begin{equation}
		{\cal P}_{AG} = (J_{H+}^d \, J_{H-}^d)(J_{H-}^d \, J_{H+}^d) 
		(1-J_{V+}^d \, J_{V-}^d)(J_{V-}^d \, J_{V+}^d)
		\label{eq:projPAG}
	\end{equation}
	and the coefficient $c_{5}$ by
	\begin{equation}
		\begin{array}{ll}
			\mbox{for \texttt{Arg}} &\qquad c_{5} = -4c_{1} + 14c_{2} \\
			\mbox{for \texttt{Ser}} &\qquad c_{5} = 12c_{2} \\
			\mbox{for \texttt{Gly}} &\qquad c_{5} = -4c_{1} + 18c_{2} \\
			\mbox{for \texttt{Ter}} &\qquad c_{5} = 6c_{1}+ 14c_{2} \\
		\end{array}
		\label{eq:coefc5}
	\end{equation}
	\item
	splitting of some doublets into singlets (one element of the singlet 
	combining to another doublet to form a triplet): \\
	\phantom{--} \texttt{Met} $\to$ \texttt{Met} + \texttt{Ile} for the EC, 
	PMC, EMC, FMC, CNC, BNC, ENC, aYNC codes; \\
	\phantom{--} \texttt{Lys} $\to$ \texttt{Lys} + \texttt{Asn} for the FMC 
	and EMC codes; \\
	\phantom{--} \texttt{Trp} $\to$ \texttt{Trp} + \texttt{Ter} for the EC, 
	CNC, BNC, aYNC codes; \\
	\phantom{--} \texttt{Trp} $\to$ \texttt{Trp} + \texttt{Cys} for the ENC 
	code; \\
	\phantom{--} \texttt{Ter} $\to$ \texttt{Tyr} + \texttt{Ter} for the FMC 
	code; \\
	Such an assignment is done through the following term in the reading 
	operator:
	\begin{equation}
		c_{6} \; {\cal P}_{XY} \; \Big( \half - J_{V,3}^{(3)} \Big) \Big( 
		\half - J_{H,3}^{(3)} \Big)
		\label{eq:readop6}
	\end{equation}
	where we use the projector ${\cal P}_{AU}$ for the splitting of the 
	\texttt{Met} doublet, ${\cal P}_{AA}$ for the \texttt{Lys} doublet, 
	${\cal P}_{UG}$ for the \texttt{Trp} doublet, and ${\cal P}_{UA}$ for 
	the \texttt{Ter} doublet.  These projectors are given by
	\begin{eqnarray}
		&& {\cal P}_{AU} = (1-J_{H+}^d \, J_{H-}^d) (J_{H-}^d \, 
		J_{H+}^d)(J_{V+}^d \, J_{V-}^d)(J_{V-}^d \, J_{V+}^d) 
		\label{eq:projPAU} \\
		&& {\cal P}_{AA} = (1-J_{H+}^d \, J_{H-}^d) (J_{H-}^d \, 
		J_{H+}^d)(1-J_{V+}^d \, J_{V-}^d)(J_{V-}^d \, J_{V+}^d) 
		\label{eq:projPAA} \\
		&& {\cal P}_{UG} = (J_{H+}^d \, J_{H-}^d) (J_{H-}^d \, 
		J_{H+}^d)(1-J_{V+}^d \, J_{V-}^d)(1-J_{V-}^d \, J_{V+}^d) 
		\label{eq:projPUG} \\
		&& {\cal P}_{UA} = (1-J_{H+}^d \, J_{H-}^d) (J_{H-}^d \, 
		J_{H+}^d)(1-J_{V+}^d \, J_{V-}^d)(1-J_{V-}^d \, J_{V+}^d)
		\label{eq:projPUA}
	\end{eqnarray}
	The coefficient $c_{6}$ takes the following values:
	\begin{equation}
		\begin{array}{ll}
			\mbox{for \texttt{Met} $\to$ \texttt{Met} + \texttt{Ile}} 
			&\qquad c_{6} = 12c_{2} \\
			\mbox{for \texttt{Lys} $\to$ \texttt{Lys} + \texttt{Asn}} 
			&\qquad c_{6} = -8c_{1} \\
			\mbox{for \texttt{Trp} $\to$ \texttt{Trp} + \texttt{Cys}} 
			&\qquad c_{6} = 12c_{2} \\
			\mbox{for \texttt{Trp} $\to$ \texttt{Trp} + \texttt{Ter}} 
			&\qquad c_{6} = 6c_{1} + 6c_{2} \\
			\mbox{for \texttt{Ter} $\to$ \texttt{Ter} + \texttt{Tyr}} 
			&\qquad c_{6} = -8c_{1} \\
		\end{array}
		\label{eq:coefc6}
	\end{equation}
	\item
	in the case of the CNC and BNC codes, the \texttt{Ter} doublet is 
	changed in \texttt{Gln} as follows: \\
	\phantom{--} \texttt{Ter} $\to$ \texttt{Gln} for the CNC code by the 
	term
	\begin{equation}
		-4c_{1} \; {\cal P}_{UA} \; \Big( \half - J_{V,3}^{(3)} \Big)
		\label{eq:readopdiv1}
	\end{equation}
	\phantom{--} \texttt{Ter} $\to$ \texttt{Ter} + \texttt{Gln} for the BNC 
	code by the term
	\begin{equation}
		-4c_{1} \; {\cal P}_{UA} \; \Big( \half - J_{V,3}^{(3)} \Big) \Big( 
		\half + J_{H,3}^{(3)} \Big)
		\label{eq:readopdiv2}
	\end{equation}
	\item
	in the case of the alternative YNC code, the last quartet \texttt{Leu} 
	is split into a triplet \texttt{Leu} coded by (CUC,CUU,CUA) and a 
	doublet \texttt{Ser} coded by (CUG).  The corresponding term in the 
	reading operator is
	\begin{equation}
		2c_{1} \; {\cal P}_{CU} \; \Big( \half - J_{V,3}^{(3)} \Big) \Big( 
		\half + J_{H,3}^{(3)} \Big)
		\label{eq:readopdiv3}
	\end{equation}
	where the projector ${\cal P}_{CU}$ is given by
	\begin{equation}
		{\cal P}_{CU} = (1-J_{H+}^d \, J_{H-}^d)(1-J_{H-}^d \, J_{H+}^d) 
		(J_{V+}^d \, J_{V-}^d)(1-J_{V-}^d \, J_{V+}^d)
		\label{eq:projPCU}
	\end{equation}
	\item
	in the case of the Yeast Mitochondrial Code, the quartet CUN codes the 
	amino-acid \texttt{Thr} rather than \texttt{Leu}.  This change is 
	achieved by multiplying the quartets term (\ref{eq:readop1}) by 
	$(1+2{\cal P}_{CU})$ for the horizontal part and by $(1-4{\cal 
	P}_{CU})$ for the vertical part.
\end{itemize}

\subsubsection{The Eukariotic Code (EC)}

The Eukariotic Code is the most important one and is often referred to as the 
universal code.  The differences between the Eukariotic Code and the prototype 
code are the following:
\begin{center}
	\begin{tabular}{cccc||cccc}
		& prototype code & EC & \quad & \quad & & prototype code & EC \\ 
		\hline
		AUG & \texttt{Met} & \texttt{Met} &&& AUA & \texttt{Met} & 
		\texttt{Ile} \\
		AGG & X & \texttt{Arg} &&& AGA & X & \texttt{Arg} \\
		UGG & \texttt{Trp} & \texttt{Trp} &&& UGA & \texttt{Trp} & 
		\texttt{Ter} \\
	\end{tabular}
\end{center}
Hence from (\ref{eq:readop5}), (\ref{eq:coefc5}), (\ref{eq:readop6}) and 
(\ref{eq:coefc6}), the reading operator for the Eukariotic Code is
\begin{eqnarray}
	{\cal R}_{EC} &=& \sfrac{4}{3} c_{1} \, C_{H} + \sfrac{4}{3} c_{2} \, 
	C_{V} - 4c_{1} \, {\cal P}_{H} \, J_{H,3} - 4c_{2} \, {\cal P}_{V} \, 
	J_{V,3} + (-8c_{1} \, {\cal P}_{D} \, + (8c_{1}+12c_{2}) \, {\cal 
	P}_{S}) \, J_{V,3} \nonumber \\
	&& + (-4c_{1} + 14c_{2}) \; {\cal P}_{AG} \; \Big( \half - 
	J_{V,3}^{(3)} \Big) \nonumber \\
	&& + \Big[ 12 c_{2} \; {\cal P}_{AU} + (6c_{1} + 6c_{2}) \; {\cal 
	P}_{UG} \Big] \Big( \half - J_{V,3}^{(3)} \Big) \Big( \half - 
	J_{H,3}^{(3)} \Big)
	\label{eq:readopEC}
\end{eqnarray}

\subsubsection{The Vertebral Mitochondrial Code (VMC)}

The Vertebral Mitochondrial Code is used in the mitochondriae of 
vertebrata.  The differences between the Vertebral Mitochondrial Code and 
the prototype code are the following:
\begin{center}
	\begin{tabular}{cccc||cccc}
		& prototype code & VMC & \quad & \quad & & prototype code & VMC \\ 
		\hline
		AGG & X & \texttt{Ter} &&& AGA & X & \texttt{Ter} \\
	\end{tabular}
\end{center}
Hence from (\ref{eq:readop5}) and (\ref{eq:coefc5}), the reading operator for 
the Vertebral Mitochondrial Code is
\begin{eqnarray}
	{\cal R}_{VMC} &=& \sfrac{4}{3} c_{1} \, C_{H} + \sfrac{4}{3} c_{2} \, 
	C_{V} - 4c_{1} \, {\cal P}_{H} \, J_{H,3} - 4c_{2} \, {\cal P}_{V} \, 
	J_{V,3} + (-8c_{1} \, {\cal P}_{D} \, + (8c_{1}+12c_{2}) \, {\cal 
	P}_{S}) \, J_{V,3} \nonumber \\
	&& + (6c_{1}+14c_{2}) \; {\cal P}_{AG} \; \Big( \half - J_{V,3}^{(3)} 
	\Big)
	\label{eq:readopVMC}
\end{eqnarray}

\subsubsection{The Yeast Mitochondrial Code (YMC)}

The Yeast Mitochondrial Code is used in the mitochondriae of yeast such as 
Saccharomyces, Candida, etc.  The differences between the Yeast 
Mitochondrial Code and the prototype code are the following:
\begin{center}
	\begin{tabular}{cccc||cccc}
		& prototype code & YMC & \quad & \quad & & prototype code & YMC \\ 
		\hline
		CUC & \texttt{Leu} & \texttt{Thr} &&& CUU & \texttt{Leu} & 
		\texttt{Thr} \\
		CUG & \texttt{Leu} & \texttt{Thr} &&& CUA & \texttt{Leu} & 
		\texttt{Thr} \\
		AGG & X & \texttt{Arg} &&& AGA & X & \texttt{Arg} \\
	\end{tabular}
\end{center}
Hence from (\ref{eq:readop5}) and (\ref{eq:coefc5}), the reading operator for 
the Yeast Mitochondrial Code is
\begin{eqnarray}
	{\cal R}_{YMC} &=& (\sfrac{4}{3} c_{1} \, C_{H} - 4c_{1} \, {\cal 
	P}_{H} \, J_{H,3})(1+2{\cal P}_{CU}) + (\sfrac{4}{3} c_{2} \, C_{V} - 
	4c_{2} \, {\cal P}_{V} \, J_{V,3})(1-4{\cal P}_{CU}) \nonumber \\
	&& + (-8c_{1} \, {\cal P}_{D} \, + (8c_{1}+12c_{2}) \, {\cal P}_{S}) \, 
	J_{V,3} + (-4c_{1}+14c_{2}) \; {\cal P}_{AG} \; \Big( \half - 
	J_{V,3}^{(3)} \Big)
	\label{eq:readopYMC}
\end{eqnarray}

\subsubsection{The Invertebrate Mitochondrial Code (IMC)}

The Invertebrate Mitochondrial Code is used in the mitochondriae of some 
arthopoda, mollusca, nematoda and insecta.  The differences between the 
Invertebrate Mitochondrial Code and the prototype code are the following:
\begin{center}
	\begin{tabular}{cccc||cccc}
		& prototype code & IMC & \quad & \quad & & prototype code & IMC \\ 
		\hline
		AGG & X & \texttt{Ser} &&& AGA & X & \texttt{Ser} \\
	\end{tabular}
\end{center}
Hence from (\ref{eq:readop5}) and (\ref{eq:coefc5}), the reading operator for 
the Invertebrate Mitochondrial Code is
\begin{eqnarray}
	{\cal R}_{IMC} &=& \sfrac{4}{3} c_{1} \, C_{H} + \sfrac{4}{3} c_{2} \, 
	C_{V} - 4c_{1} \, {\cal P}_{H} \, J_{H,3} - 4c_{2} \, {\cal P}_{V} \, 
	J_{V,3} + (-8c_{1} \, {\cal P}_{D} \, + (8c_{1}+12c_{2}) \, {\cal 
	P}_{S}) \, J_{V,3} \nonumber \\
	&& + 12c_{2} \; {\cal P}_{AG} \; \Big( \half - J_{V,3}^{(3)} \Big)
	\label{eq:readopIMC}
\end{eqnarray}

\subsubsection{The Protozoan Mitochondrial and Mycoplasma Code (PMC)}

The Protozoan Mitochondrial and Mycoplasma Code is used in the 
mitochondriae of some protozoa (leishmania, paramecia, trypanosoma, etc.)  
and for many fungi.  The differences between the Protozoan Mitochondrial 
and Mycoplasma Code and the prototype code are the following:
\begin{center}
	\begin{tabular}{cccc||cccc}
		& prototype code & PMC & \quad & \quad & & prototype code & PMC \\ 
		\hline
		AUG & \texttt{Met} & \texttt{Met} &&& AUA & \texttt{Met} & 
		\texttt{Ile} \\
		AGG & X & \texttt{Arg} &&& AGA & X & \texttt{Arg} \\
	\end{tabular}
\end{center}
Hence from (\ref{eq:readop5}), (\ref{eq:coefc5}), (\ref{eq:readop6}) and 
(\ref{eq:coefc6}), the reading operator for the Protozoan Mitochondrial Code 
is
\begin{eqnarray}
	{\cal R}_{PMC} &=& \sfrac{4}{3} c_{1} \, C_{H} + \sfrac{4}{3} c_{2} \, 
	C_{V} - 4c_{1} \, {\cal P}_{H} \, J_{H,3} - 4c_{2} \, {\cal P}_{V} \, 
	J_{V,3} + (-8c_{1} \, {\cal P}_{D} \, + (8c_{1}+12c_{2}) \, {\cal 
	P}_{S}) \, J_{V,3} \nonumber \\
	&& + (-4c_{1}+14c_{2}) \; {\cal P}_{AG} \; \Big( \half - J_{V,3}^{(3)} 
	\Big) + 12 c_{2} \; {\cal P}_{AU} \; \Big( \half - J_{V,3}^{(3)} \Big) 
	\Big( \half - J_{H,3}^{(3)} \Big)
	\label{eq:readopPMC}
\end{eqnarray}

\subsubsection{The Echinoderm Mitochondrial Code (EMC)}

The Echinoderm Mitochondrial Code is used in the mitochondriae of some 
asterozoa and echinozoa.  The differences between the Echinoderm 
Mitochondrial Code and the prototype code are the following:
\begin{center}
	\begin{tabular}{cccc||cccc}
		& prototype code & EMC & \quad & \quad & & prototype code & EMC \\ 
		\hline
		AUG & \texttt{Met} & \texttt{Met} &&& AUA & \texttt{Met} & 
		\texttt{Ile} \\
		AGG & X & \texttt{Ser} &&& AGA & X & \texttt{Ser} \\
		AAG & \texttt{Lys} & \texttt{Lys} &&& AAA & \texttt{Lys} & 
		\texttt{Asn} \\
	\end{tabular}
\end{center}
Hence from (\ref{eq:readop5}), (\ref{eq:coefc5}), (\ref{eq:readop6}) and 
(\ref{eq:coefc6}), the reading operator for the Echinoderm Mitochondrial Code 
is
\begin{eqnarray}
	{\cal R}_{EMC} &=& \sfrac{4}{3} c_{1} \, C_{H} + \sfrac{4}{3} c_{2} \, 
	C_{V} - 4c_{1} \, {\cal P}_{H} \, J_{H,3} - 4c_{2} \, {\cal P}_{V} \, 
	J_{V,3} + (-8c_{1} \, {\cal P}_{D} \, + (8c_{1}+12c_{2}) \, {\cal 
	P}_{S}) \, J_{V,3} \nonumber \\
	&& + 12 c_{2} \; {\cal P}_{AG} \; \Big( \half - J_{V,3}^{(3)} \Big) + 
	\Big[ 12 c_{2} \; {\cal P}_{AU} - 8 c_{1} \; {\cal P}_{AA} \Big] \Big( 
	\half - J_{V,3}^{(3)} \Big) \Big( \half - J_{H,3}^{(3)} \Big)
	\label{eq:readopEMC}
\end{eqnarray}

\subsubsection{The Ascidian Mitochondrial Code (AMC)}

The Ascidian Mitochondrial Code is used in the mitochondriae of some 
ascidiacea.  The differences between the Ascidian Mitochondrial Code and 
the prototype code are the following:
\begin{center}
	\begin{tabular}{cccc||cccc}
		& prototype code & AMC & \quad & \quad & & prototype code & AMC \\ 
		\hline
		AGG & X & \texttt{Gly} &&& AGA & X & \texttt{Gly} \\
	\end{tabular}
\end{center}
Hence from (\ref{eq:readop5}) and (\ref{eq:coefc5}), the reading operator for 
the Ascidian Mitochondrial Code is
\begin{eqnarray}
	{\cal R}_{AMC} &=& \sfrac{4}{3} c_{1} \, C_{H} + \sfrac{4}{3} c_{2} \, 
	C_{V} - 4c_{1} \, {\cal P}_{H} \, J_{H,3} - 4c_{2} \, {\cal P}_{V} \, 
	J_{V,3} + (-8c_{1} \, {\cal P}_{D} \, + (8c_{1}+12c_{2}) \, {\cal 
	P}_{S}) \, J_{V,3} \nonumber \\
	&& + (-4c_{1}+18c_{2}) \; {\cal P}_{AG} \; \Big( \half - J_{V,3}^{(3)} 
	\Big)
	\label{eq:readopAMC}
\end{eqnarray}

\subsubsection{The Flatworm Mitochondrial Code (FMC)}

The Flatworm Mitochondrial Code is used in the mitochondriae of the 
flatworms.  The differences between the Flatworm Mitochondrial Code and the 
prototype code are the following:
\begin{center}
	\begin{tabular}{cccc||cccc}
		& prototype code & FMC & \quad & \quad & & prototype code & FMC \\ 
		\hline
		UAG & \texttt{Ter} & \texttt{Ter} &&& UAA & \texttt{Ter} & 
		\texttt{Tyr} \\
		AUG & \texttt{Met} & \texttt{Met} &&& AUA & \texttt{Met} & 
		\texttt{Ile} \\
		AGG & X & \texttt{Ser} &&& AGA & X & \texttt{Ser} \\
		AAG & \texttt{Lys} & \texttt{Lys} &&& AAA & \texttt{Lys} & 
		\texttt{Asn} \\
	\end{tabular}
\end{center}
Hence from (\ref{eq:readop5}), (\ref{eq:coefc5}), (\ref{eq:readop6}) and 
(\ref{eq:coefc6}), the reading operator for the Flatworm Mitochondrial Code is
\begin{eqnarray}
	{\cal R}_{FMC} &=& \sfrac{4}{3} c_{1} \, C_{H} + \sfrac{4}{3} c_{2} \, 
	C_{V} - 4c_{1} \, {\cal P}_{H} \, J_{H,3} - 4c_{2} \, {\cal P}_{V} \, 
	J_{V,3} + (-8c_{1} \, {\cal P}_{D} \, + (8c_{1}+12c_{2}) \, {\cal 
	P}_{S}) \, J_{V,3} \nonumber \\
	&& + 12 c_{2} \; {\cal P}_{AG} \; \Big( \half - J_{V,3}^{(3)} \Big) + 
	\Big[ 12 c_{2} \; {\cal P}_{AU} - 8 c_{1} \; {\cal P}_{AA} - 8 c_{1} \; 
	{\cal P}_{UA} \Big] \Big( \half - J_{V,3}^{(3)} \Big) \Big( \half - 
	J_{H,3}^{(3)} \Big) \nonumber \\
	\label{eq:readopFMC}
\end{eqnarray}

\subsubsection{The Ciliate Nuclear Code (CNC)}

The Ciliate Nuclear Code is used in the nuclei of some ciliata, 
dasyclasaceae and diplomonadida.  The differences between the Ciliate 
Nuclear Code and the prototype code are the following:
\begin{center}
	\begin{tabular}{cccc||cccc}
		& prototype code & CNC & \quad & \quad & & prototype code & CNC \\ 
		\hline
		UGG & \texttt{Trp} & \texttt{Trp} &&& UGA & \texttt{Trp} & 
		\texttt{Ter} \\
		UAG & \texttt{Ter} & \texttt{Gln} &&& UAA & \texttt{Ter} & 
		\texttt{Gln} \\
		AUG & \texttt{Met} & \texttt{Met} &&& AUA & \texttt{Met} & 
		\texttt{Ile} \\
		AGG & X & \texttt{Arg} &&& AGA & X & \texttt{Arg} \\
	\end{tabular}
\end{center}
Hence from (\ref{eq:readop5}), (\ref{eq:coefc5}), (\ref{eq:readop6}), 
(\ref{eq:coefc6}) and (\ref{eq:readopdiv1}), the reading operator for the 
Ciliate Nuclear Code is
\begin{eqnarray}
	{\cal R}_{CNC} &=& \sfrac{4}{3} c_{1} \, C_{H} + \sfrac{4}{3} c_{2} \, 
	C_{V} - 4c_{1} \, {\cal P}_{H} \, J_{H,3} - 4c_{2} \, {\cal P}_{V} \, 
	J_{V,3} + (-8c_{1} \, {\cal P}_{D} \, + (8c_{1}+12c_{2}) \, {\cal 
	P}_{S}) \, J_{V,3} \nonumber \\
	&& + \Big[ (-4c_{1}+14c_{2}) \; {\cal P}_{AG} - 4c_{1} \; {\cal P}_{UA} 
	\Big] \Big( \half - J_{V,3}^{(3)} \Big) \nonumber \\
	&& + \Big[ 12 c_{2} \; {\cal P}_{AU} + (6 c_{1} + 6 c_{2}) \; {\cal 
	P}_{UG} \Big] \Big( \half - J_{V,3}^{(3)} \Big) \Big( \half - 
	J_{H,3}^{(3)} \Big)
	\label{eq:readopCNC}
\end{eqnarray}

\subsubsection{The Blepharisma Nuclear Code (BNC)}

The Blepharisma Nuclear Code is used in the nuclei of the blepharisma 
(ciliata) (note that this code is very close to the CNC which is used for 
the ciliata).  The differences between the Blepharisma Nuclear Code and the 
prototype code are the following:
\begin{center}
	\begin{tabular}{cccc||cccc}
		& prototype code & BNC & \quad & \quad & & prototype code & BNC \\ 
		\hline
		UGG & \texttt{Trp} & \texttt{Trp} &&& UGA & \texttt{Trp} & 
		\texttt{Ter} \\
		UAG & \texttt{Ter} & \texttt{Gln} &&& UAA & \texttt{Ter} & 
		\texttt{Ter} \\
		AUG & \texttt{Met} & \texttt{Met} &&& AUA & \texttt{Met} & 
		\texttt{Ile} \\
		AGG & X & \texttt{Arg} &&& AGA & X & \texttt{Arg} \\
	\end{tabular}
\end{center}
Hence from (\ref{eq:readop5}), (\ref{eq:coefc5}), (\ref{eq:readop6}), 
(\ref{eq:coefc6}) and (\ref{eq:readopdiv2}), the reading operator for the 
Blepharisma Nuclear Code is
\begin{eqnarray}
	{\cal R}_{BNC} &=& \sfrac{4}{3} c_{1} \, C_{H} + \sfrac{4}{3} c_{2} \, 
	C_{V} - 4c_{1} \, {\cal P}_{H} \, J_{H,3} - 4c_{2} \, {\cal P}_{V} \, 
	J_{V,3} + (-8c_{1} \, {\cal P}_{D} \, + (8c_{1}+12c_{2}) \, {\cal 
	P}_{S}) \, J_{V,3} \nonumber \\
	&& + (-4c_{1}+14c_{2}) \; {\cal P}_{AG} \; \Big( \half - J_{V,3}^{(3)} 
	\Big) - 4c_{1} \; {\cal P}_{UA} \; \Big( \half - J_{V,3}^{(3)} \Big) 
	\Big( \half + J_{H,3}^{(3)} \Big) \nonumber \\
	&& + \Big[ 12 c_{2} \; {\cal P}_{AU} + (6 c_{1} + 6 c_{2}) \; {\cal 
	P}_{UG} \Big] \Big( \half - J_{V,3}^{(3)} \Big) \Big( \half - 
	J_{H,3}^{(3)} \Big)
	\label{eq:readopBNC}
\end{eqnarray}

\subsubsection{The Euplotid Nuclear Code (ENC)}

The Euplotid Nuclear Code is used in the nuclei of the euplotidae 
(ciliata).  The differences between the Euplotid Nuclear Code and the 
prototype code are the following:
\begin{center}
	\begin{tabular}{cccc||cccc}
		& prototype code & ENC & \quad & \quad & & prototype code & ENC \\ 
		\hline
		UGG & \texttt{Trp} & \texttt{Trp} &&& UGA & \texttt{Trp} & 
		\texttt{Cys} \\
		AUG & \texttt{Met} & \texttt{Met} &&& AUA & \texttt{Met} & 
		\texttt{Ile} \\
		AGG & X & \texttt{Arg} &&& AGA & X & \texttt{Arg} \\
	\end{tabular}
\end{center}
Hence from (\ref{eq:readop5}), (\ref{eq:coefc5}), (\ref{eq:readop6}) and 
(\ref{eq:coefc6}), the reading operator for the Euplotid Nuclear Code is
\begin{eqnarray}
	{\cal R}_{ENC} &=& \sfrac{4}{3} c_{1} \, C_{H} + \sfrac{4}{3} c_{2} \, 
	C_{V} - 4c_{1} \, {\cal P}_{H} \, J_{H,3} - 4c_{2} \, {\cal P}_{V} \, 
	J_{V,3} + (-8c_{1} \, {\cal P}_{D} \, + (8c_{1}+12c_{2}) \, {\cal 
	P}_{S}) \, J_{V,3} \nonumber \\
	&& + (-4c_{1}+14c_{2}) \; {\cal P}_{AG} \; \Big( \half - J_{V,3}^{(3)} 
	\Big) + 12 c_{2} \; ({\cal P}_{AU} + {\cal P}_{UG} ) \Big( \half - 
	J_{V,3}^{(3)} \Big) \Big( \half - J_{H,3}^{(3)} \Big)
	\label{eq:readopENC}
\end{eqnarray}

\subsubsection{The alternative Yeast Nuclear Code (alt.  YNC)}

The alternative Yeast Nuclear Code is used in the nuclei of some yeast 
(essentially many candidae).  The differences between the alternative Yeast 
Nuclear Code and the prototype code are the following:
\begin{center}
	\begin{tabular}{cccc||cccc}
		& prototype code & alt.  YNC & \quad & \quad & & prototype code & 
		alt.  YNC \\ \hline
		CUG & \texttt{Leu} & \texttt{Ser} &&& CUA & \texttt{Leu} & 
		\texttt{Leu} \\
		UGG & \texttt{Trp} & \texttt{Trp} &&& UGA & \texttt{Trp} & 
		\texttt{Ter} \\
		AUG & \texttt{Met} & \texttt{Met} &&& AUA & \texttt{Met} & 
		\texttt{Ile} \\
		AGG & X & \texttt{Arg} &&& AGA & X & \texttt{Arg} \\
	\end{tabular}
\end{center}
Hence from (\ref{eq:readop5}), (\ref{eq:coefc5}), (\ref{eq:readop6}), 
(\ref{eq:coefc6}) and (\ref{eq:readopdiv3}), the reading operator for the 
alternative Yeast Nuclear Code is
\begin{eqnarray}
	{\cal R}_{aYNC} &=& \sfrac{4}{3} c_{1} \, C_{H} + \sfrac{4}{3} c_{2} \, 
	C_{V} - 4c_{1} \, {\cal P}_{H} \, J_{H,3} - 4c_{2} \, {\cal P}_{V} \, 
	J_{V,3} + (-8c_{1} \, {\cal P}_{D} \, + (8c_{1}+12c_{2}) \, {\cal 
	P}_{S}) \, J_{V,3} \nonumber \\
	&& + (-4c_{1}+14c_{2}) \; {\cal P}_{AG} \; \Big( \half - J_{V,3}^{(3)} 
	\Big) + 2c_{1} \; {\cal P}_{CU} \; \Big( \half - J_{V,3}^{(3)} \Big) 
	\Big( \half + J_{H,3}^{(3)} \Big) \nonumber \\
	&& + \Big[ (6 c_{1} + 6 c_{2}) \; {\cal P}_{UG} + 12 c_{2} \; {\cal 
	P}_{AU} \Big] \Big( \half - J_{V,3}^{(3)} \Big) \Big( \half - 
	J_{H,3}^{(3)} \Big)
	\label{eq:readopaYNC}
\end{eqnarray}

\subsection{Reading values for the amino-acids}

We have therefore constructed reading operators for the genetic codes 
specified above, starting from a prototype code that emphasizes the 
quartet/doublet structure of the different codes.  The different reading 
operators are such that they give the same value for a given amino-acid, 
whatever the code under consideration.  Finally, we get the following 
eigenvalues of the reading operators for the amino-acids (after a rescaling, 
setting $c \equiv c_{1}/c_{2}$):
\begin{equation}
	\begin{array}{c|r||c|r||c|r}
		\mbox{a.a.} & \phantom{-} \mbox{value of ${\cal R}$} \phantom{-} &
		\mbox{a.a.} & \phantom{-} \mbox{value of ${\cal R}$}\phantom{-}  &
		\mbox{a.a.} & \phantom{-} \mbox{value of ${\cal R}$} \phantom{-} \\
		\hline
		\bigg. \mbox{Ala} & - c + 3 \phantom{--} &
		\bigg. \mbox{Gly} & - c + 5 \phantom{--} &
		\bigg. \mbox{Pro} & - c - 1 \phantom{--} \\
		\bigg. \mbox{Arg} & - c + 1 \phantom{--} & 
		\bigg. \mbox{His} & -3 c + 1 \phantom{--} &
		\bigg. \mbox{Ser} & 3 c - 1 \phantom{--} \\
		\bigg. \mbox{Asn} & 9 c + 5 \phantom{--} &
		\bigg. \mbox{Ile} & 5 c + 9 \phantom{--} &
		\bigg. \mbox{Thr} & 3 c + 3 \phantom{--} \\
		\bigg. \mbox{Asp} & 5 c + 5 \phantom{--} &
		\bigg. \mbox{Leu} & c - 1 \phantom{--} & 
		\bigg. \mbox{Trp} & 3 c - 5 \phantom{--} \\
		\bigg. \mbox{Cys} & 3 c + 7 \phantom{--} &
		\bigg. \mbox{Lys} & 17 c + 5 \phantom{--} & 
		\bigg. \mbox{Tyr} & c + 1 \phantom{--} \\
		\bigg. \mbox{Gln} & 5 c + 1 \phantom{--} &
		\bigg. \mbox{Met} & 5 c - 3 \phantom{--} & 
		\bigg. \mbox{Val} & c + 3 \phantom{--} \\
		\bigg. \mbox{Glu} & 13 c + 5 \phantom{--} &
		\bigg. \mbox{Phe} & -7 c - 1 \phantom{--} & 
		\bigg. \mbox{Ter} & 9 c + 1 \phantom{--} \\
	\end{array}
\end{equation}
Remark that the reading operators ${\cal R}(c)$ can be used for any real 
value of $c$, except those conferring the same eigenvalue to codons 
relative to two different amino-acids.  These forbidden values are the 
following: $-7$, $-5$, $-4$, $-3$, $-\sfrac{5}{2}$, $-\sfrac{7}{3}$, $-2$, 
$-\sfrac{5}{3}$, $-\thalf$, $-\sfrac{4}{3}$, $-1$, $-\sfrac{5}{6}$, 
$-\sfrac{4}{5}$, $-\sfrac{3}{4}$, $-\sfrac{5}{7}$, $-\sfrac{2}{3}$, 
$-\sfrac{3}{5}$, $-\half$, $-\sfrac{3}{7}$, $-\sfrac{2}{5}$, 
$-\sfrac{3}{8}$, $-\sfrac{1}{3}$, $-\sfrac{3}{10}$, $-\sfrac{2}{7}$, 
$-\sfrac{1}{4}$, $-\sfrac{2}{9}$, $-\sfrac{1}{5}$, $-\sfrac{1}{6}$, 
$-\sfrac{1}{7}$, $-\sfrac{1}{8}$, $-\sfrac{1}{9}$, $0$, $\sfrac{1}{7}$, 
$\sfrac{1}{6}$, $\sfrac{1}{5}$, $\sfrac{1}{4}$, $\sfrac{1}{3}$, 
$\sfrac{2}{5}$, $\half$, $\sfrac{2}{3}$, $1$, $\sfrac{4}{3}$, $\thalf$, 
$2$, $\sfrac{5}{2}$, $3$, $4$, $5$.

\bigskip

At this point, let us emphasize the specific properties of our model.  To 
each nucleotide are assigned specific quantum numbers characterizing its 
purine/pyrimidine origin and involving the complementary rule.  Then 
ordered sequences of bases can be constructed and characterized in this 
framework.  Ordered sequences of three bases have been just above examined 
and the correspondence codon/amino-acid represented by the reading operator 
${\cal R}$.  Finally let us remark that the coefficients $c_{i}$, which 
above have been taken as constants, can more generally be considered as 
functions of some external variables (biological, physical and chemical 
environment, time, etc.).  In this way it is possible to explain the 
observed discrepancy in the correspondence codons/amino-acid in biological 
species under stress conditions (in vitro).  In this scheme the evolution 
process of genetic code can also be discussed.  However, we believe that a 
better understanding of the reasons of the evolution, i.e. which kind of 
optimization process takes place, has still to be acquired.

\section{Physical properties of the dinucleotides}
\label{fits}

The model we have at hand, with nucleotides characterized by quantum 
numbers, is well adapted to elaborate formulae expressing biophysical 
properties.  A particularly interesting quantity is the free energy 
released by base pairing in double stranded RNA. The data are not provided 
for a doublet of nucleotides, with one item in each strand, but for a pair 
of nucleotides, for ex.  CG, lying on one strand and coupled with another 
pair, i.e. GC on the second strand ; note also that the direction on a 
strand being perfectly defined, the release of energy for the doublet 
sequence CG on the first strand running from $5'$ to $3'$ related to the 
doublet GC on the complementary strand running from $3'$ to $5'$, will be 
different to the one related to the doublet GC, itself associated to CG. It 
appears clear that such quantities involve pairs of nucleotides, and that 
naturally ordered crystal bases obtained from tensor product of two 
representations are adapted for such a calculation.

We will also consider two other quantities involving again pairs of 
nucleotides, namely the relative hydrophilicity $R_{f}$ and hydrophobicity 
$R_{x}$ of dinucleosides.

Before presenting our results, let us mention that fits for the same 
biophysical properties can be found in a recent preprint \cite{BJ} where 
polynomials in 4 or 6 coordinates in the 64 codon space are constructed.  
In their approach, the authors associate two coordinates $(d,m)$ to each 
nucleotide of any codon, as follows: $A = (-1,0)$, $C = (0,-1)$, $G = 
(0,1)$, $U = (1,0)$, labelling in this way each codon with 6 numbers.  The 
above labelling of the nucleotides is related to our labels Eq.  
(\ref{eq:gc1}) in the following way:
\begin{displaymath}
\begin{array}{c|c|c}
	& d & m \\
	\hline
	C & J_{V,3} - J_{H,3} & - (J_{V,3} + J_{H,3}) \\
	U & J_{V,3} - J_{H,3} & - (J_{V,3} + J_{H,3}) \\
	G & J_{V,3} + J_{H,3} & J_{V,3} - J_{H,3} \\
	A & J_{V,3} + J_{H,3} & J_{V,3} - J_{H,3} \\
\end{array}
\end{displaymath}
Therefore the labels $(d,m)$ just correspond up to a sign for the pyrimidine 
(resp.  purine) to the antidiagonal and diagonal (resp.  diagonal and 
antidiagonal) ${\cal U}_{q \to 0}(sl(2))$.

In the following we compare our results with those of \cite{BJ}.

\subsubsection*{Free energy}

In \cite{FSS1} we have fitted the experimental data with a four-parameter 
operator.  Here we fit the more recent data \cite{MSZT} with a 
two-parameter operator obtained from the one used in \cite{FSS1} by setting 
two parameters to zero:
\begin{equation}
	\Delta G^0_{37} = \alpha_{0} + \alpha_{1} (C_{H}+C_{V}) J_{3H}^d
	\label{eq:dg}
\end{equation}
Using a least-squares fit, one finds for the coefficients $\alpha_{i}$:
\begin{equation}
	\alpha_{0} = -2.14 \,, \;\; \alpha_{1} = -0.295
	\label{eq:dgbestfit}
\end{equation}
The standard deviation of the two-parameter fit (\ref{eq:dgbestfit}) is 
found to be equal to $0.149$, which is to be compared to the standard 
deviation $0.16$ of the four-parameter fit of ref.  \cite{BJ}.  The 
experimental and fitted values of the free energies $\Delta G^0_{37}$ of 
the dinucleotides are displayed in Table \ref{tbl:dg37}.
\begin{table}[htbp]
\centering
\[
\begin{array}{cc}
	\left[ \mbox{CA} \;\; \raisebox{-2mm}{\shortstack{$-2.1\;\;$\\$-2.14$}} 
	\right] & \left[
	\begin{array}{lll} 
		\mbox{CG} \;\; \raisebox{-2mm}{\shortstack{$-2.4\;\;$\\$-2.73$}} & 
		\mbox{UG} \;\; \raisebox{-2mm}{\shortstack{$-2.1\;\;$\\$-2.14$}} & 
		\mbox{UA} \;\; \raisebox{-2mm}{\shortstack{$-1.3\;\;$\\$-1.55$}} \\
	\end{array}
	\right] \\
	\\
	\left[
	\begin{array}{c} 
		\mbox{CU} \;\; \raisebox{-2mm}{\shortstack{$-2.1\;\;$\\$-2.14$}} \\ 
		\\
		\mbox{GU} \;\; \raisebox{-2mm}{\shortstack{$-2.2\;\;$\\$-2.14$}} \\ 
		\\
		\mbox{GA} \;\; \raisebox{-2mm}{\shortstack{$-2.4\;\;$\\$-2.14$}} \\
	\end{array} 
	\right] & \left[ 
	\begin{array}{lll} 
		\mbox{CC} \;\; \raisebox{-2mm}{\shortstack{$-3.3\;\;$\\$-3.32$}} & 
		\mbox{UC} \;\; \raisebox{-2mm}{\shortstack{$-2.4\;\;$\\$-2.14$}} & 
		\mbox{UU} \;\; \raisebox{-2mm}{\shortstack{$-0.9\;\;$\\$-0.96$}} \\
		&& \\
		\mbox{GC} \;\; \raisebox{-2mm}{\shortstack{$-3.4\;\;$\\$-3.32$}} & 
		\mbox{AC} \;\; \raisebox{-2mm}{\shortstack{$-2.2\;\;$\\$-2.14$}} & 
		\mbox{AU} \;\; \raisebox{-2mm}{\shortstack{$-1.1\;\;$\\$-0.96$}} \\
		&& \\
		\mbox{GG} \;\; \raisebox{-2mm}{\shortstack{$-3.3\;\;$\\$-3.32$}} & 
		\mbox{AG} \;\; \raisebox{-2mm}{\shortstack{$-2.1\;\;$\\$-2.14$}} & 
		\mbox{AA} \;\; \raisebox{-2mm}{\shortstack{$-0.9\;\;$\\$-0.96$}} \\
	\end{array} 
	\right]
\end{array}
\]
\caption{Dinucleotides free energies $\Delta G^0_{37}$.}
The upper (resp. lower) values are the experimental (resp. fitted) values.
\label{tbl:dg37}
\end{table}

\subsubsection*{Hydrophilicity}

We fit the values of the relative hydrophilicity $R_{f}$ of the 16 
dinucleoside monophosphates \cite{WL} with the following four-parameter 
operator:
\begin{equation}
	R_{f} = \alpha_{0} + \alpha_{1} C_{V} + \alpha_{2} J_{3V}^d + 
	\alpha_{3} \sum_{i=1,2} (J_{3H}^i + J_{3V}^i)(J_{3H}^i + J_{3V}^i - 1)
	\label{eq:RF}
\end{equation}
(the last term in $\alpha_{3}$ is equal to 4 for AA, to 2 for CA, GA, UA 
and zero for the other dinucleotides).  \\
Using a least-squares fit, one finds for the coefficients $\alpha_{i}$:
\begin{equation}
	\alpha_{0} = 0.135 \,, \;\; \alpha_{1} = 0.036 \,, \;\; \alpha_{2} = 
	0.147 \,, \;\; \alpha_{3} = -0.016
	\label{eq:RFbestfit}
\end{equation}
The standard deviation of the four-parameter fit (\ref{eq:RFbestfit}) is 
found to be equal to $0.027$, which is to be compared to the standard 
deviation $0.033$ of the six-parameter fit of ref.  \cite{BJ}.  The 
experimental and fitted values of the hydrophilicity $R_{f}$ of the 
dinucleosides are displayed in Table \ref{tbl:rf}.
\begin{table}[htbp]
\centering
\[
\begin{array}{cc}
	\left[ \mbox{CA} \;\; \raisebox{-2mm}{\shortstack{0.083\\0.103}} 
	\right] & \left[
	\begin{array}{lll} 
		\mbox{CG} \;\; \raisebox{-2mm}{\shortstack{0.146\\0.135}} & 
		\mbox{UG} \;\; \raisebox{-2mm}{\shortstack{0.160\\0.135}} & 
		\mbox{UA} \;\; \raisebox{-2mm}{\shortstack{0.090\\0.103}} \\
	\end{array}
	\right] \\
	\\
	\left[
	\begin{array}{c} 
		\mbox{CU} \;\; \raisebox{-2mm}{\shortstack{0.359\\0.354}} \\ 
		\\
		\mbox{GU} \;\; \raisebox{-2mm}{\shortstack{0.224\\0.207}} \\ 
		\\
		\mbox{GA} \;\; \raisebox{-2mm}{\shortstack{0.035\\0.028}} \\
	\end{array} 
	\right] & \left[ 
	\begin{array}{lll} 
		\mbox{CC} \;\; \raisebox{-2mm}{\shortstack{0.349\\0.354}} & 
		\mbox{UC} \;\; \raisebox{-2mm}{\shortstack{0.378\\0.354}} & 
		\mbox{UU} \;\; \raisebox{-2mm}{\shortstack{0.389\\0.354}} \\
		&& \\
		\mbox{GC} \;\; \raisebox{-2mm}{\shortstack{0.193\\0.207}} & 
		\mbox{AC} \;\; \raisebox{-2mm}{\shortstack{0.118\\0.175}} & 
		\mbox{AU} \;\; \raisebox{-2mm}{\shortstack{0.112\\0.175}} \\
		&& \\
		\mbox{GG} \;\; \raisebox{-2mm}{\shortstack{0.065\\0.060}} & 
		\mbox{AG} \;\; \raisebox{-2mm}{\shortstack{0.048\\0.028}} & 
		\mbox{AA} \;\; \raisebox{-2mm}{\shortstack{0.023\\-0.004}} \\
	\end{array} 
	\right]
\end{array}
\]
\caption{Dinucleosides relative hydrophilicities $R_{f}$.}
The upper (resp. lower) values are the experimental (resp. fitted) values.
\label{tbl:rf}
\end{table}

\subsubsection*{Hydrophobicity}

We fit the values of the relative hydrophobicity $R_{x}$ of the 16 
dinucleoside monophosphates as reported in \cite{J} with the following 
four-parameter operator:
\begin{equation}
	R_{x} = \alpha_{0} + \alpha_{1} J_{3V}^d + \alpha_{2} J_{3H}^d + 
	\alpha_{3} [(J_{3H}^1 + J_{3V}^1)^2 + (J_{3H}^2 + J_{3V}^2)^2]
	\label{eq:RX}
\end{equation}
(the last term in $\alpha_{3}$ is equal to 2 for AA, AC, CA and CC, to 2 
for AU, AG, UA, UC, GC, GA, CU and CG and zero for UU, UG, GU, GG).  \\
Using a least-squares fit, one finds for the coefficients $\alpha_{i}$:
\begin{equation}
	\alpha_{0} = 0.294 \,, \;\; \alpha_{1} = -0.240 \,, \;\; \alpha_{2} = 
	-0.105 \,, \;\; \alpha_{3} = 0.136
	\label{eq:RXbestfitAA}
\end{equation}
Using a least-squares fit without the dinucleoside AA, one finds new 
coefficients $\alpha_{i}$, which lead to better values of $R_{x}$ for the 
remaining dinucleosides:
\begin{equation}
	\alpha_{0} = 0.309 \,, \;\; \alpha_{1} = -0.203 \,, \;\; \alpha_{2} = 
	-0.068 \,, \;\; \alpha_{3} = 0.099
	\label{eq:RXbestfitsansAA}
\end{equation}
\begin{table}[htbp]
\centering
\[
\begin{array}{cc}
	\left[ \mbox{CA} \;\; \raisebox{-2mm}{\shortstack{0.494\\0.507}} 
	\right] & \left[
	\begin{array}{lll} 
		\mbox{CG} \;\; \raisebox{-2mm}{\shortstack{0.326\\0.340}} & 
		\mbox{UG} \;\; \raisebox{-2mm}{\shortstack{0.291\\0.309}} & 
		\mbox{UA} \;\; \raisebox{-2mm}{\shortstack{0.441\\0.476}} \\
	\end{array}
	\right] \\
	\\
	\left[
	\begin{array}{c} 
		\mbox{CU} \;\; \raisebox{-2mm}{\shortstack{0.218\\0.205}} \\ 
		\\
		\mbox{GU} \;\; \raisebox{-2mm}{\shortstack{0.291\\0.309}} \\ 
		\\
		\mbox{GA} \;\; \raisebox{-2mm}{\shortstack{0.660\\0.611}} \\
	\end{array} 
	\right] & \left[ 
	\begin{array}{lll} 
		\mbox{CC} \;\; \raisebox{-2mm}{\shortstack{0.244\\0.236}} & 
		\mbox{UC} \;\; \raisebox{-2mm}{\shortstack{0.218\\0.205}} & 
		\mbox{UU} \;\; \raisebox{-2mm}{\shortstack{0.194\\0.174}} \\
		&& \\
		\mbox{GC} \;\; \raisebox{-2mm}{\shortstack{0.326\\0.340}} & 
		\mbox{AC} \;\; \raisebox{-2mm}{\shortstack{0.494\\0.507}} & 
		\mbox{AU} \;\; \raisebox{-2mm}{\shortstack{0.441\\0.476}} \\
		&& \\
		\mbox{GG} \;\; \raisebox{-2mm}{\shortstack{0.436\\0.444}} & 
		\mbox{AG} \;\; \raisebox{-2mm}{\shortstack{0.660\\0.611}} & 
		\mbox{AA} \;\; \raisebox{-2mm}{\shortstack{1\\0.778}} \\
	\end{array} 
	\right]
\end{array}
\]
\caption{Dinucleosides relative hydrophobicities $R_{x}$.}
The upper (resp. lower) values are the experimental (resp. fitted) values.
\label{tbl:rx}
\end{table}
The standard deviation of the four-parameter fit (\ref{eq:RXbestfitAA}) is 
equal to $0.049$, which is the same of the four-parameter fit of ref.  
\cite{BJ}.  Using the fit (\ref{eq:RXbestfitsansAA}), the standard 
deviation becomes $0.074$ (including the value for AA) or $0.024$ 
(excluding the value for AA).  For this last case, the standard deviation 
of ref.  \cite{BJ} is still equal to $0.031$.  The experimental and fitted 
values (second fit) of the relative hydrophobicity $R_{x}$ of the 
dinucleosides are displayed in Table \ref{tbl:rx}.

\section{Universal behaviour of ratios of codon usage frequency}
\label{correl}

In the following the labels $X,J,Z,K$ represent any of the 4 bases $C,U,G,A$.  
Let $XJZ$ be a codon in a given multiplet, say $m_{i}$, encoding an a.a., say 
$A_{i}$.  We define the probability of usage of the codon $XJZ$ as the ratio 
between the frequency of usage $n_{Z}$ of the codon $XJZ$ in the biosynthesis 
of $A_{i}$ and the total number $n$ of synthesized $A_{i}$, i.e.  as the 
relative codon frequency, in the limit of \emph{very large $n$}.

It is natural to assume that the usage frequency of a codon in a multiplet 
is connected to its probability of usage $P(XJZ \to A_{i})$.  We define 
\cite{FSS2} the \emph{branching ratio} $B_{ZK}$ as
\begin{equation}
	B_{ZK} = \frac{P(XJZ \to A_{i})}{P(XJK \to A_{i})}
	\label{eq:brrat}
\end{equation}
where $XJK$ is another codon belonging to the same multiplet $m_{i}$.  It 
is reasonable to argue that in the limit of very large number of codons, 
for a fixed biological species and amino-acid, the branching ratio depends 
essentially on the properties of the codon.  In our model this means that 
in this limit $B_{ZK}$ is a function, depending on the type of the 
multiplet, on the \emph{quantum numbers} of the codons $XJZ$ and $XJK$, 
i.e. on the labels $J_{\alpha}, J_{\alpha,3}$, where $\alpha = H$ or $V$, 
and on an other set of quantum labels leaving out the degeneracy on 
$J_{\alpha}$; in Table \ref{tablerep} different irreducible representations 
with the same values of $J_{\alpha}$ are distinguished by an upper label.  

\medskip

We have put in evidence a correlation in the codon usage frequency for the 
quartets and the quartet subpart of the sextets, i.e. the codons in a 
sextet differing only for the third codon, for the vertebrates in 
\cite{FSS2} and for biological species belonging to the vertebrates, 
invertebrates, plants and fungi in \cite{CFSS}, and we have shown that 
these correlations fit well in our model with the assumed dependence on 
$B_{ZK}$. Here we remark that for thirteen biological species belonging to 
the vertebrate class, with a statistics of codons larger than 95,000 (see 
Table \ref{tutu}), the ratio of
\begin{equation}
	\frac{B_{AG}}{B_{UC}} = \frac{B_{AU}}{B_{GC}} = \frac{P(XJA \to 
	A_{i})}{P(XJG \to A_{i})} \; \frac{P(XJC \to A_{i})}{P(XJU \to A_{i})}
\end{equation}
for quartets and the quartet subpart of the sextets has a behaviour 
independent of the specific biological species.  Moreover, for the same 
amino-acids for which we have remarked correlations, the values of the 
ratio $B_{AG}/B_{UC}$ are almost the same (see Table \ref{ratio}).  We show 
that these behaviour and correlations find a nice explanation in our model.  
In Tables \ref{ratioAG} and \ref{ratioUC}, we report respectively the 
values of the branching ratios $B_{AG}$ and $B_{UC}$ as computed from the 
database \cite{data} (release of February 2000) and in Table \ref{ratio} 
the ratio of these quantities.  The average values $\langle B_{AG}/B_{UC} 
\rangle$, the standard deviations $\sigma$ and the ratios $\sigma/\langle 
B_{AG}/B_{UC} \rangle$ are displayed in the following table:
\begin{displaymath}
\begin{array}{|l||r|r|r|r|r|r|r|r|}
	\hline
	& \mathrm{Pro} & \mathrm{Ala} & \mathrm{Thr} & \mathrm{Ser} & 
	\mathrm{Gly} & \mathrm{Val} & \mathrm{Leu} & \mathrm{Arg} \\
	\hline
	\langle B_{AG}/B_{UC} \rangle & 2.50 & 2.84 & 3.30 & 2.67 & 2.21 & 0.33 
	& 0.26 & 1.32 \\
	\sigma & 0.46 & 0.53 & 0.56 & 0.35 & 0.30 & 0.04 & 0.03 & 0.14 \\
	\sigma/\langle B_{AG}/B_{UC} \rangle & 0.19 & 0.19 & 0.17 & 0.13 & 0.14 
	& 0.13 & 0.10 & 0.11 \\
	\hline
\end{array}
\end{displaymath}
The above behaviour can be easily understood considering a dependence on 
$B_{ZK}$ not only on the irreducible representations to which the codons 
$XJZ$ and $XJK$ appearing in the numerator and the denominator belong, but 
also on the specific states denoting these codons, and refining the 
factorized form of \cite{FSS2} as
\begin{equation}
	B_{ZK} = F_{ZK}(IR(XJZ);IR(XJK)) \; \frac{G_{H}(b.s.;J_{H,3}(XJZ)) \, 
	G_{V}(b.s.;J_{V,3}(XJZ))}{G_{H}(b.s.;J_{H,3}(XJK)) \, 
	G_{V}(b.s.;J_{H,3}(XJK))}
	\label{eq:bzk}
\end{equation}
where we have denoted by $b.s.$ the biological species, by $IR(XJZ)$ and 
$J_{\alpha,3}(XJZ)$ the irreducible representation to which the codon $XJZ$ 
belong (see Table \ref{tablerep}), and the value of the third component of 
the $\alpha$-spin of the state $XJZ$.  Note that we have still neglected 
the dependence on the type of the biosynthetized amino-acid.  \emph{The 
ratio $B_{AG}/B_{UC}$ using Eq.  (\ref{eq:bzk}), is no more depending on 
the biological species but only on the value of the irreducible 
representations of the codons}.  Moreover, for \texttt{Pro}, \texttt{Ala}, 
\texttt{Thr}, \texttt{Ser}, (resp.  \texttt{Val} and \texttt{Leu}), the 
irreducible representations appearing in the $F$ functions are the same as 
can be seen from Table \ref{irreps}, so we expect the same value for the 
ratio, which is indeed the case (see above Table), the value of 
$B_{AG}/B_{UC}$ for the first four amino-acids (resp.  for the last two 
amino-acids) lying in the range $2.90 \pm 15 \%$ (resp.  $0.30 \pm 15 \%$).  
These values should be compared with the value 1.32 for \texttt{Arg} and 
2.21 for \texttt{Gly}.

Let us end this section by the following remark.  From the above table, one 
might be tempted to consider the value of the ratio $B_{AG}/B_{UC}$ for 
\texttt{Gly} of the same order of magnitude as the ones for \texttt{Pro}, 
\texttt{Ala}, \texttt{Thr}, \texttt{Ser}.  Then one distinguishes, 
following this ratio, three groups of codons quartets: the one associated 
to the five just mentioned amino-acids, another one relative to 
\texttt{Val} and \texttt{Leu}, and a last one with \texttt{Arg}.  Now, let 
us look at the dinucleotide pairs constituting the first two nucleotides in 
a codon in the light of our results of sect.  \ref{model}: the pairs CC, 
GC, AC, UC and GG relative to \texttt{Pro}, \texttt{Ala}, \texttt{Thr}, 
\texttt{Ser}, and \texttt{Gly} respectively belong to the representation 
(1,1) of ${\cal U}_{q \to 0}(sl(2) \oplus sl(2))$; the states GU and CU 
relative to \texttt{Val} and \texttt{Leu} respectively belong to the 
representation (0,1); finally CG relative to \texttt{Arg} also lies in a 
different representation (1,0).

\section {Mutations in the genetic code}
\label{mutation}

In this section, we present a mathematical framework to describe the 
single-base deletions in the genetic code.  In \cite{JK} starting from the 
observation that the single-base deletions in DNA, which occur far more 
frequently that single base additions, take place in the opposite site to a 
purine \textbf{R}, (\textbf{R = G, A}) i.e. a pyrimidine \textbf{Y} 
(\textbf{Y = C, U/T}) is deleted, arguments have been presented to explain 
why the Stop codons have the structure they have, see Table \ref{tablerep}.  
We refer to the paper for more details and for references to the biological 
literature on the subject and we recall here just the main ideas and 
conclusions of \cite{JK}.  The starting point is the observed fact that 
deletions occur more frequently in the following sequences: \textbf{YR}, 
\textbf{TTR}, \textbf{YTG} and \textbf{TR}.  In ref.  \cite{JK} all these 
sequences have been refined as \textbf{YTRV}, (\textbf{V = C, A, G}).  
Starting from the structure of this dangerous sequence and using the 
complementarity property, an analysis shows that four codons -- 
\textbf{TAA}, \textbf{TAG}, \textbf{TTA}, \textbf{CTA} -- are both 
potential deletion site codons and reverse-complementary potential site 
codons.  As a mutation at the end of a protein chain just implies the 
addition of further peptides, the authors conclude that the assignment of 
codons \textbf{TAA} and \textbf{TAG} as Stop codons minimizes the possible 
deleterious effects of deletion.  Indeed the codon usage frequency of the 
dangerous codon \textbf{CTA}, as it can be seen from fig.  (5) of 
\cite{FSS2} and from fig.  (2) of \cite{CFSS}, is very low.  An analysis of 
the codon usage frequency exhibits an analogous behaviour for the codon 
\textbf{TTA}.

The mechanism by which the above specified sequences are preferred in the 
deletion process is unclear.  In the following we will present a 
mathematical scheme in which these properties can be settled.  Let us 
recall that the Wigner-Eckart theorem, has been extended to the quantum 
algebra $U_q(sl(n))$, and recently in \cite{MS} to the case of $U_{q \to 
0}(sl(2))$.

In \cite{MS} ($q \to 0$)-tensor operators have been introduced, called 
crystal tensor operators, which transform as
\begin{equation}
	J_{3}(\tau_{m}^{j}) \equiv m \tau_{m}^{j} \quad J_{\pm} \, 
	(\tau_{m}^{j}) \equiv \tau_{m \pm 1}^{j}
\end{equation}

Clearly, if $|m| > j$ then $\tau_{m}^{j}$  has to be considered vanishing.

The ($q \to 0$)-Wigner-Eckart theorem can be written ($ j_{1} \ge j $)
\begin{eqnarray}
	\tau_{m}^{j} \, \vert j_{1} m_{1} \rangle &=& (-1)^{2j} \, \sum_{\alpha 
	= 0}^{2j} \, \langle j_{1} + j - \alpha \Vert \tau^{j} \Vert j_{1} 
	\rangle \, | j_{1} + j - \alpha, m_{1} + m \rangle \, \nonumber \\
	&& (\delta_{m_{1}, j_{1} - \alpha} + \delta_{-m, j - \alpha} - 
	\delta_{m_{1}, j_{1} - \alpha} \, \delta_{-m, j - \alpha})
\end{eqnarray}

The ($q \to 0$)-Wigner-Eckart theorem has the peculiar feature that the 
selection rules do not depend only on the rank of the tensor operator and 
on the initial state, but in a crucial way from the specific component of 
the tensor in consideration.  The tensor product of two irreducible 
representations in the crystal basis is not commutative (see sect.  
\ref{model}), therefore one has to specify which is the first 
representation.  In the following, as in \cite{MS}, the crystal tensor 
operator has to be considered as the first one.

Let us also remark the following peculiar property of crystal basis which 
will be used in the following.  We specify it only for the case we are 
interested in, but it is a completely general property.

An ordered sequence, or chain, of $n$ nucleotides is a state belonging to 
an irreducible representation of $U_{q \to 0}((sl(2) \oplus sl(2))$ 
appearing in the $n$-fold product of the fundamental irreducible 
representation $(1/2,1/2)$.  Moreover the same property holds for any 
subsequence of $m$ ($m < n$) nucleotides.  We can mimick the deletion of a 
$N$ nucleotide in a generic position of a coding sequence by a local 
annihilation operator of the $N$ nucleotide.  In order to take into account 
the observed fact that the deletion of the nucleotide depends on the nature 
of the neighboring nucleotides, we require the annihilation operator to 
behave as a defined crystal tensor operator under $U_{q \to 0}(sl(2))_{V}$ 
or $U_{q \to 0}(sl(2))_{H}$ or both.  In our mathematical description we 
have to specify the action of the annihilation operator on a chain of 
nucleotides.  If we assume that the annihilation of the $N$ nucleotide 
behaves e.g. as a spinor crystal operator for the $U_{q \to 0}(sl(2))_{V}$, 
we have to require that the deletion of the $N$ nucleotide from the initial 
chain of $K$ nucleotides, described by the state $\vert J_{i}, M_{i}; 
\Omega_{i} \rangle$, leading to the final chain of $K-1$ nucleotides, 
described by the state $\vert J_{f}, M_{f}; \Omega_{f} \rangle$, is 
compatible with the ($q \to 0$)-Wigner-Eckart theorem prescription for the 
action of the definite crystal spinor operator between the initial state 
$\vert J_{i}, M_{i}; \Omega_{i} \rangle$ and the final state $\vert J_{f}, 
M_{f}; \Omega_{f} \rangle$, where we have denoted by $\Omega$ the set of 
all the labels necessary to identify completely the state.  As we shall 
see, this is far from being trivial and will put constraints on the type of 
nucleotides surrounding the nucleotide $N$.  We have to specify which chain 
has to be considered in order to study the action of the crystal tensor 
operator.  It seems reasonable to take into account chains formed by $K=2$ 
and 3 nucleotides starting from $N$ in the sense of the reading of the 
codon sequence.  So we are defining on the chain the action of a 
``matrioska'' crystal tensor operator.  We assume:

\noindent \underline{ Assumption} : The biological mechanism responsible 
for the deletion of a pyrimidine \textbf{C} (resp.  \textbf{U} ) in a 
sequence can be schematized by a local crystal tensor operator 
$\tau_{-1/2}^{1/2}$ for $U_{q \to 0}(sl(2)_{V})$ and $\tau_{-1/2}^{1/2}$ 
(resp.  $\tau_{1/2}^{1/2}$ ) for $U_{q \to 0}(sl(2)_{H})$, which transforms 
the state \textbf{YX} (resp. \textbf{YXZ}) into the state \textbf{X} 
(resp. \textbf{XZ}), \textbf{X}, \textbf{Z} being any nucleotide.

By ``local crystal tensor operator'' we mean an operator which, in the 
sequence of RNA, acts on the $K$-chain ($K = 2,3$) starting with 
\textbf{Y}, deleting the pyrimidine, according to the selection rules 
imposed by the assumed type of the crystal tensor.

Let us point out that, differently to ref.  \cite{JK}, where the DNA 
sequence was analyzed, we consider the transcripted RNA sequence and the 
deletion in the trascription of a \textbf{Y}.

There are 8 possible cases (we denote the initial and final states with the 
notation of sect.  \ref{model} and by A (resp.  F) the allowed (resp.  
forbidden) transition).  We analyze the deletion of a \textbf{C} (on the 
left) and of an \textbf{U} (on the right).
\begin{displaymath}
\begin{tabular}{|c|c|c|}
	\multicolumn{3}{c}{$_{\Bigg.}$ Action of $ \tau_{-1/2,H}^{1/2} \oplus 
	\tau_{-1/2,V}^{1/2}$} \\
	\hline
	\multicolumn{2}{|c}{$_{\Big.}^{\Big.} (1,1) \to (\half,\half)$} & \\
	\hline
	\textbf{CC} & \textbf{C} & F--F \\
	\hline
	\multicolumn{2}{|c}{$_{\Big.}^{\Big.} (0,1) \to (\half,\half)$} & \\
	\hline
	\textbf{CU} & \textbf{U} & A--F \\
	\hline
	\multicolumn{2}{|c}{$_{\Big.}^{\Big.} (1,0) \to (\half,\half)$} & \\
	\hline
	\textbf{CG} & \textbf{G} & F--A \\
	\hline
	\multicolumn{2}{|c}{$_{\Big.}^{\Big.} (0,0) \to (\half,\half)$} & \\
	\hline
	\textbf{CA} & \textbf{A} & A--A \\
	\hline
\end{tabular}
\qquad \qquad \qquad \qquad
\begin{tabular}{|c|c|c|}
	\multicolumn{3}{c}{$_{\Bigg.}$ Action of $ \tau_{1/2,H}^{1/2} \oplus 
	\tau_{-1/2,V}^{1/2}$} \\
	\hline
	\multicolumn{2}{|c}{$_{\Big.}^{\Big.} (1,1) \to (\half,\half)$} & \\
	\hline
	\textbf{UC} & \textbf{C} & A--F \\
	\textbf{UU} & \textbf{U} & A--F \\
	\hline
	\multicolumn{2}{|c}{$_{\Big.}^{\Big.} (1,0) \to (\half,\half)$} & \\
	\hline
	\textbf{UG} & \textbf{G} & A--A \\
	\textbf{UA} & \textbf{A} & A--A \\
	\hline
\end{tabular}
\end{displaymath}
So for the transition for the state of dinucleotide to one nucleotide 
state, from the assumed nature of the crystal tensor operator, it follows 
that a pyrimidine can be deleted if followed by a purine.  Now let us 
consider what happens if we consider the transition from a trinucleotide to 
a dinucletide state.  Using the previous result we consider only the state 
in which a purine is in second position so we have to consider 16 cases:
\begin{displaymath}
\begin{tabular}{|c|c|c|}
	\multicolumn{3}{c}{$_{\Bigg.}$ Action of $ \tau_{-1/2,H}^{1/2} \oplus 
	\tau_{-1/2,V}^{1/2}$} \\
	\hline
	\multicolumn{2}{|c}{$_{\Big.}^{\Big.} (\half,\half)^4 \to (1,1)$} & \\
	\hline
	\textbf{CAC} & \textbf{AC} & A--A \\
	\textbf{CAU} & \textbf{AU} & A--A \\
	\textbf{CAA} & \textbf{AA} & A--A \\
	\textbf{CAG} & \textbf{AG} & A--A \\
	\hline
	\multicolumn{2}{|c}{$_{\Big.}^{\Big.} (\thalf,\half)^2 \to (1,1)$} & \\
	\hline
	\textbf{CGC} & \textbf{GC} & F--A \\
	\textbf{CGG} & \textbf{GG} & F--A \\
	\hline
	\multicolumn{2}{|c}{$_{\Big.}^{\Big.} (\half,\half)^2 \to (0,1)$} & \\
	\hline
	\textbf{CGU} & \textbf{GU} & F--A \\
	\textbf{CGA} & \textbf{GA} & F--A \\
	\hline
\end{tabular}
\qquad \qquad \qquad \qquad
\begin{tabular}{|c|c|c|}
	\multicolumn{3}{c}{$_{\Bigg.}$ Action of $ \tau_{1/2,H}^{1/2} \oplus 
	\tau_{-1/2,V}^{1/2}$} \\
	\hline
	\multicolumn{2}{|c}{$_{\Big.}^{\Big.} (\thalf,\half)^2 \to (1,1)$} & \\
	\hline
	\textbf{UAC} & \textbf{AC} & A--A \\
	\textbf{UAU} & \textbf{AU} & A--A \\
	\textbf{UAA} & \textbf{AA} & A--A \\
	\textbf{UAG} & \textbf{AG} & A--A \\
	\hline
	\multicolumn{2}{|c}{$_{\Big.}^{\Big.} (\thalf,\half)^2 \to (1,1)$} & \\
	\hline
	\textbf{UGC} & \textbf{GC} & A--A \\
	\textbf{UGG} & \textbf{GG} & A--A \\
	\hline
	\multicolumn{2}{|c}{$_{\Big.}^{\Big.} (\half,\half)^2 \to (0,1)$} & \\
	\hline
	\textbf{UGU} & \textbf{GU} & A--A \\
	\textbf{UGA} & \textbf{GA} & A--A \\
	\hline
\end{tabular}
\end{displaymath}
So, from the assumed nature of the crystal tensor operator, the transition 
from a trinucleotide to a dinucleotide state is horizontally forbidden for 
the deletion of a \textbf{C} if the second nucleotide is a \textbf{G}.
 
Let us note that we have made the simplified assuption that the transitions 
depend only on the values of $J_{\alpha}, J_{\alpha,3}$ of the initial 
and final state.  

Moreover, both to take into account the data of \cite{JK} and to check that 
the results are not very sensible to the choice of the initial state, we 
consider the deletion of a purine in second position in a four-nucleotide 
state and impose that the process may take place only if the initial and 
final state can be connected by a spinor crystal operator 
$\tau_{-1/2,H}^{1/2} \oplus \tau_{-1/2,V}^{1/2}$ for the deletion of 
\textbf{C} or $\tau_{1/2,H}^{1/2} \oplus \tau_{-1/2,V}^{1/2}$ for the 
deletion of \textbf{U}.

As the two pyrimidines differ by their value of $J_{H,3}$, the constraints 
imposed by the tensor operator $\tau_{\pm 1/2,H}^{1/2}$ are weaker than 
those imposed by the tensor operator $\tau_{-1/2,V}^{1/2}$.

In Appendix (in sect.  \ref{model}) we have reported all the irreducible 
representations arising by the $4$-fold ($3$-fold) tensor product of the 
fundamental representation.  A detailed analysis shows that only the 
following deletions may happen (we report all the transitions that are 
allowed at least once):
\begin{displaymath}
\begin{tabular}{|c|c|c|}
	\multicolumn{3}{c}{$_{\Bigg.}$ Action of $\tau_{-1/2,H}^{1/2} \oplus 
	\tau_{-1/2,V}^{1/2}$} \\
	\hline
	\multicolumn{2}{|c}{$_{\Big.}^{\Big.} (2,1)^3 \to (\thalf,\thalf)$} & \\
	\hline
	\textbf{GCGC} & \textbf{GGC} & F--A \\
	\textbf{ACGC} & \textbf{AGC} & F--A \\
	\textbf{GCGG} & \textbf{GGG} & F--A \\
	\textbf{ACGG} & \textbf{AGG} & F--A \\
	\hline
	\multicolumn{2}{|c}{$_{\Big.}^{\Big.} (2,0)^2 \to (\thalf,\half)^2$} & \\
	\hline
	\textbf{CCGG} & \textbf{CGG} & F--A \\
	\textbf{UCGG} & \textbf{UGG} & F--A \\
	\hline
	\multicolumn{2}{|c}{$_{\Big.}^{\Big.} (1,1)^7 \to (\half,\thalf)^1$} & \\
	\hline
	\textbf{GCGU} & \textbf{GGU} & F--A \\
	\textbf{ACGU} & \textbf{AGU} & F--A \\
	\textbf{GCGA} & \textbf{GGA} & F--A \\
	\textbf{ACGA} & \textbf{AGA} & F--A \\
	\hline
	\multicolumn{2}{|c}{$_{\Big.}^{\Big.} (1,0)^4 \to (\half,\half)^2$} & \\
	\hline
	\textbf{CCGA} & \textbf{CGA} & F--A \\
	\textbf{UCGA} & \textbf{UGA} & F--A \\
	\hline
	\multicolumn{2}{|c}{$_{\Big.}^{\Big.} (1,1)^9 \to (\half,\thalf)^2$} & \\
	\hline
	\textbf{GCAC} & \textbf{GAC} & F--A \\
	\textbf{GCAG} & \textbf{GAG} & F--A \\
	\hline
	\multicolumn{2}{|c}{$_{\Big.}^{\Big.} (1,0)^6 \to (\half,\half)^4$} & \\
	\hline
	\textbf{CCAG} & \textbf{CAG} & F--A \\
	\hline
	\multicolumn{2}{|c}{$_{\Big.}^{\Big.} (1,1)^9 \to (\thalf,\thalf)$} & \\
	\hline
	\textbf{ACAC} & \textbf{AAC} & A--A \\
	\textbf{ACAU} & \textbf{AAU} & A--A \\
	\textbf{ACAG} & \textbf{AAG} & A--A \\
	\textbf{ACAA} & \textbf{AAA} & A--A \\
	\hline
	\multicolumn{2}{|c}{$_{\Big.}^{\Big.} (1,0)^6 \to (\thalf,\half)^2$} & \\
	\hline
	\textbf{UCAG} & \textbf{UAG} & A--A \\
	\textbf{UCAA} & \textbf{UAA} & A--A \\
	\hline
\end{tabular}
\qquad \qquad \qquad \qquad
\begin{tabular}{|c|c|c|}
	\multicolumn{3}{c}{$_{\Bigg.}$ Action of $ \tau_{-1/2,H}^{1/2} \oplus 
	\tau_{-1/2,V}^{1/2}$} \\
	\hline
	\multicolumn{2}{|c}{$_{\Big.}^{\Big.} (1,2)^3 \to (\thalf,\thalf)$} & \\
	\hline
	\textbf{UCUC} & \textbf{UUC} & A--F \\
	\textbf{UCUU} & \textbf{UUU} & A--F \\
	\textbf{ACUC} & \textbf{AUC} & A--F \\
	\textbf{ACUU} & \textbf{AUU} & A--F \\
	\hline
	\multicolumn{2}{|c}{$_{\Big.}^{\Big.} (0,2)^2 \to (\half,\thalf)^2$} & \\
	\hline
	\textbf{CCUU} & \textbf{CUU} & A--F \\
	\textbf{GCUU} & \textbf{GUU} & A--F \\
	\hline
	\multicolumn{2}{|c}{$_{\Big.}^{\Big.} (1,1)^8 \to (\thalf,\half)^1$} & \\
	\hline
	\textbf{UCUG} & \textbf{UUG} & A--F \\
	\textbf{UCUA} & \textbf{UUA} & A--F \\
	\textbf{ACUG} & \textbf{AUG} & A--F \\
	\textbf{ACUA} & \textbf{AUA} & A--F \\
	\hline
	\multicolumn{2}{|c}{$_{\Big.}^{\Big.} (0,1)^5 \to (\half,\half)^3$} & \\
	\hline
	\textbf{CCUA} & \textbf{CUA} & A--F \\
	\textbf{GCUA} & \textbf{GUA} & A--F \\
	\hline
	\multicolumn{2}{|c}{$_{\Big.}^{\Big.} (1,1)^9 \to (\thalf,\half)^2$} & \\
	\hline
	\textbf{UCAC} & \textbf{UAC} & A--F \\
	\textbf{UCAU} & \textbf{UAU} & A--F \\
	\hline
	\multicolumn{2}{|c}{$_{\Big.}^{\Big.} (0,1)^6 \to (\half,\half)^4$} & \\
	\hline
	\textbf{CCAU} & \textbf{CAU} & A--F \\
	\hline
	\multicolumn{2}{|c}{$_{\Big.}^{\Big.} (0,0)^4 \to (\half,\half)^4$} & \\
	\hline
	\textbf{CCAA} & \textbf{CAA} & A--A \\
	\hline
	\multicolumn{2}{|c}{$_{\Big.}^{\Big.} (0,1)^6 \to (\half,\thalf)^2$} & \\
	\hline
	\textbf{GCAU} & \textbf{GAU} & A--A \\
	\textbf{GCAA} & \textbf{GAA} & A--A \\
	\hline
	\multicolumn{3}{c}{} \\
	\multicolumn{3}{c}{} \\
	\multicolumn{3}{c}{} \\
\end{tabular}
\end{displaymath}
So we remark:
\begin{itemize}
	\item 
	The deletion of \textbf{C}, allowed or horizontally forbidden, may 
	happen only if it is followed by a purine.  In the allowed cases, it 
	must be followed by the nucleotide \textbf{A}, in agreement with the 
	observed data.
	\item 
	A nucleotide \textbf{A} before the deleted nucleotide \textbf{C} 
	appears only in the transition $(1,1)^9 \to (\thalf,\thalf)$.  This 
	feature is present in the observed data with a very low occurrence, 
	which in our language would mean that the matrix element of $\tau$ 
	between these two irreducible representations is small.
\end{itemize}    

Now we consider the case of deletion of \textbf{U}.  A detailed analysis 
shows that only the following deletions may happen:
\begin{displaymath}
\begin{tabular}{|c|c|c|}
	\multicolumn{3}{c}{$_{\Bigg.}$ Action of $ \tau_{1/2,H}^{1/2} \oplus 
	\tau_{-1/2,V}^{1/2}$} \\
	\hline
	\multicolumn{2}{|c}{$_{\Big.}^{\Big.} (1,2)^1 \to (\half,\thalf)^2$} & \\
	\hline
	\textbf{CUUC} & \textbf{CUC} & A--F \\
	\textbf{CUUU} & \textbf{CUU} & A--F \\
	\textbf{GUUC} & \textbf{GUC} & A--F \\
	\textbf{GUUU} & \textbf{GUU} & A--F \\
	\hline
	\multicolumn{2}{|c}{$_{\Big.}^{\Big.} (1,2)^2 \to (\thalf,\thalf)$} & \\
	\hline
	\textbf{CUCC} & \textbf{CCC} & A--F \\
	\textbf{GUCC} & \textbf{GCC} & A--F \\
	\hline
	\multicolumn{2}{|c}{$_{\Big.}^{\Big.} (1,1)^3 \to (\half,\half)^4$} & \\
	\hline
	\textbf{CUAC} & \textbf{CAC} & A--F \\
	\textbf{CUAU} & \textbf{CAU} & A--F \\
	\hline
	\multicolumn{2}{|c}{$_{\Big.}^{\Big.} (1,1)^6 \to (\half,\half)^3$} & \\
	\hline
	\textbf{UUCA} & \textbf{UCA} & A--F \\
	\textbf{AUCA} & \textbf{ACA} & A--F \\
	\hline
	\multicolumn{2}{|c}{$_{\Big.}^{\Big.} (2,1)^2 \to (\thalf,\half)^1$} & \\
	\hline
	\textbf{UUCG} & \textbf{UCG} & A--F \\
	\textbf{UUUG} & \textbf{UUG} & A--F \\
	\textbf{UUUA} & \textbf{UUA} & A--F \\
	\textbf{AUCG} & \textbf{ACG} & A--F \\
	\textbf{AUUG} & \textbf{AUG} & A--F \\
	\textbf{AUUA} & \textbf{AUA} & A--F \\
	\hline
	\multicolumn{2}{|c}{$_{\Big.}^{\Big.} (2,1)^3 \to (\thalf,\half)^2$} & \\
	\hline
	\textbf{UUGC} & \textbf{UGC} & A--F \\
	\textbf{UUAC} & \textbf{UAC} & A--F \\
	\textbf{UUAU} & \textbf{UAU} & A--F \\
	\hline
	\multicolumn{2}{|c}{$_{\Big.}^{\Big.} (1,1)^7 \to (\half,\half)^3$} & \\
	\hline
	\textbf{UUGU} & \textbf{UGU} & A--F \\
	\hline
	\multicolumn{2}{|c}{$_{\Big.}^{\Big.} (0,1)^4 \to (\half,\half)^2$} & \\
	\hline
	\textbf{CUGU} & \textbf{CGU} & A--F \\
	\hline
\end{tabular}
\qquad \qquad \qquad \qquad
\begin{tabular}{|c|c|c|}
	\multicolumn{3}{c}{$_{\Bigg.}$ Action of $ \tau_{1/2,H}^{1/2} \oplus 
	\tau_{-1/2,V}^{1/2}$} \\
	\hline
	\multicolumn{2}{|c}{$_{\Big.}^{\Big.} (1,1)^2 \to (\half,\half)^3$} & \\
	\hline
	\textbf{CUUG} & \textbf{CUG} & A--F \\
	\textbf{GUUG} & \textbf{GUG} & A--F \\
	\textbf{CUUA} & \textbf{CUA} & A--F \\
	\textbf{GUUA} & \textbf{GUA} & A--F \\
	\hline
	\multicolumn{2}{|c}{$_{\Big.}^{\Big.} (1,1)^2 \to (\thalf,\half)^1$} & \\
	\hline
	\textbf{CUCG} & \textbf{CCG} & A--F \\
	\textbf{GUCG} & \textbf{GCG} & A--F \\
	\hline
	\multicolumn{2}{|c}{$_{\Big.}^{\Big.} (1,2)^2 \to (\half,\thalf)^1$} & \\
	\hline
	\textbf{UUCU} & \textbf{UCU} & A--F \\
	\textbf{AUCU} & \textbf{ACU} & A--F \\
	\hline
	\multicolumn{2}{|c}{$_{\Big.}^{\Big.} (0,2)^1 \to (\half,\thalf)^1$} & \\
	\hline
	\textbf{CUCU} & \textbf{CCU} & A--F \\
	\textbf{GUCU} & \textbf{GCU} & A--F \\
	\hline
	\multicolumn{2}{|c}{$_{\Big.}^{\Big.} (2,2) \to (\thalf,\thalf)$} & \\
	\hline
	\textbf{UUCC} & \textbf{UCC} & A--F \\
	\textbf{UUUC} & \textbf{UUC} & A--F \\
	\textbf{UUUU} & \textbf{UUU} & A--F \\
	\textbf{AUCC} & \textbf{ACC} & A--F \\
	\textbf{AUUC} & \textbf{AUC} & A--F \\
	\textbf{AUUU} & \textbf{AUU} & A--F \\
	\hline
	\multicolumn{2}{|c}{$_{\Big.}^{\Big.} (0,1)^3 \to (\half,\half)^1$} & \\
	\hline
	\textbf{CUCA} & \textbf{CCA} & A--F \\
	\textbf{GUCA} & \textbf{GCA} & A--F \\
	\hline
	\multicolumn{2}{|c}{$_{\Big.}^{\Big.} (1,1)^3 \to (\thalf,\half)^2$} & \\
	\hline
	\textbf{CUGC} & \textbf{CGC} & A--F \\
	\hline
	\multicolumn{3}{c}{$_{\Big.}^{\Big.}$} \\
	\multicolumn{3}{c}{} \\
	\multicolumn{3}{c}{} \\
\end{tabular}
\end{displaymath}

\begin{displaymath}
\begin{tabular}{|c|c|c|}
	\multicolumn{3}{c}{$_{\Bigg.}$ Action of $ \tau_{1/2,H}^{1/2} \oplus 
	\tau_{-1/2,V}^{1/2}$} \\
	\hline
	\multicolumn{2}{|c}{$_{\Big.}^{\Big.} (1,1)^7 \to (\half,\thalf)^1$} & \\
	\hline
	\textbf{AUGU} & \textbf{AGU} & A--A \\
	\textbf{AUGA} & \textbf{AGA} & A--A \\
	\hline
	\multicolumn{2}{|c}{$_{\Big.}^{\Big.} (0,1)^4 \to (\half,\thalf)^1$} & \\
	\hline
	\textbf{GUGU} & \textbf{GGU} & A--A \\
	\textbf{GUGA} & \textbf{GGA} & A--A \\
	\hline
	\multicolumn{2}{|c}{$_{\Big.}^{\Big.} (1,1)^3 \to (\half,\thalf)^2$} & \\
	\hline
	\textbf{GUAC} & \textbf{GAC} & A--A \\
	\textbf{GUAU} & \textbf{GAU} & A--A \\
	\textbf{GUAG} & \textbf{GAG} & A--A \\
	\textbf{GUAA} & \textbf{GAA} & A--A \\
	\hline
	\multicolumn{2}{|c}{$_{\Big.}^{\Big.} (2,0)^2 \to (\thalf,\half)^2$} & \\
	\hline
	\textbf{UUGG} & \textbf{UGG} & A--A \\
	\textbf{UUAG} & \textbf{UAG} & A--A \\
	\textbf{UUAA} & \textbf{UAA} & A--A \\
	\hline
	\multicolumn{2}{|c}{$_{\Big.}^{\Big.} (1,0)^2 \to (\thalf,\half)^2$} & \\
	\hline
	\textbf{CUGG} & \textbf{CGG} & A--A \\
	\hline
\end{tabular}
\qquad \qquad \qquad \qquad
\begin{tabular}{|c|c|c|}
	\multicolumn{3}{c}{$_{\Bigg.}$ Action of $ \tau_{1/2,H}^{1/2} \oplus 
	\tau_{-1/2,V}^{1/2}$} \\
	\hline
	\multicolumn{2}{|c}{$_{\Big.}^{\Big.} (1,1)^3 \to (\thalf,\thalf)$} & \\
	\hline
	\textbf{GUGC} & \textbf{GGC} & A--A \\
	\textbf{GUGG} & \textbf{GGG} & A--A \\
	\hline
	\multicolumn{2}{|c}{$_{\Big.}^{\Big.} (1,0)^2 \to (\half,\half)^4$} & \\
	\hline
	\textbf{CUAG} & \textbf{CAG} & A--A \\
	\textbf{CUAA} & \textbf{CAA} & A--A \\
	\hline
	\multicolumn{2}{|c}{$_{\Big.}^{\Big.} (2,1)^3 \to (\thalf,\thalf)$} & \\
	\hline
	\textbf{AUGC} & \textbf{AGC} & A--A \\
	\textbf{AUAC} & \textbf{AAC} & A--A \\
	\textbf{AUAU} & \textbf{AAU} & A--A \\
	\textbf{AUGG} & \textbf{AGG} & A--A \\
	\textbf{AUAG} & \textbf{AAG} & A--A \\
	\textbf{AUAA} & \textbf{AAA} & A--A \\
	\hline
	\multicolumn{2}{|c}{$_{\Big.}^{\Big.} (0,0)^2 \to (\half,\half)^2$} & \\
	\hline
	\textbf{CUGA} & \textbf{CGA} & A--A \\
	\hline
	\multicolumn{3}{c}{$_{\Big.}^{\Big.}$} \\
	\multicolumn{3}{c}{} \\
\end{tabular}
\end{displaymath}
So we remark:
\begin{itemize}
	\item
	The deletion of \textbf{U} may happen only if it is followed by 
	\textbf{A} or by \textbf{G}.  In the observed data only \textbf{A} is 
	considered; however in \cite{JK} the reported deletion of \textbf{U} 
	are about 1/4 with respect to the reported deletion of \textbf{C}.  So 
	our modelisation just foresees a different environment for the deletion 
	of \textbf{U} and \textbf{C}.
	\item
	The last nucleotide in the four-nucleotide sequence in which the 
	deletion occurs may be any nucleotide, but the case in which it is a 
	purine seems more frequent than the case in which it is a pyrimidine.
	\item
	There are no transition which are only horizontally forbidden.
\end{itemize}

In conclusion, both from considering the transitions on the $K$-chains 
($K=2,3$) to the $(K-1)$-chains or the transition from the four-nucleotide 
states to the three-nucleotide states under the action of the crystal 
tensor operators, we deduce that the deletion of a pyrimidine may happen if 
it is followed by a purine.  In particular, for the deletion of C the 
preferred purine is the adenine A, whilst for the deletion of U also the 
guanine G may appear.  This makes a difference between the two cases and it 
would be extremely interesting to see if more accurate data may confirm 
this asymmetry.  Moreover the next following nucleotide may be of any type 
but there is indication that a purine is preferred.  So our mathematical 
scheme explains the main features of the observed data \cite{JK}.  A more 
quantitative analysis should require higher statistics in the experimental 
data.

\section{Recent theoretical approaches: a comparison}
\label{compare}

The use of continuous symmetries in the genetic code has been considered by 
different teams these recent years\footnote{See section ``Symmetry 
techniques in Biological Systems'' in Proc.  XXII Int.  Coll.  on Group 
Theoretical Methods in Physics, pp.  142-165.}.  It appears of some 
importance to summarize each of these approaches, and to make clear how the 
model we propose differ from them.
  
In 1993, an underlying symmetry based on a continuous group has been 
proposed \cite{HH}.  More precisely, considering the eukaryotic code, the 
authors tried to answer the following question: is it possible to determine 
a Lie algebra ${\cal G}$ carrying a 64-dimensional irreducible 
representation $R$ and admitting a subalgebra ${\cal H}$ such that the 
decomposition of $R$ into irreducible multiplets under ${\cal H}$ gives 
exactly the 21 different multiplets, the different codons in each of the 
first 20 multiplets being associated to the same amino-acid, the last 
multiplet containing the stop codons ?  They proposed as starting symmetry 
the symplectic algebra $sp(6)$, which indeed admits an irreducible 
representation of dimension 64, equal to the number of different codons, 
with the successive breakings:
\begin{equation}
	sp(6) \supset sp(4) \oplus su(2) \supset su(2) \oplus su(2) \oplus 
	su(2) \supset su(2) \oplus U(1) \oplus su(2) \supset su(2) \oplus U(1) 
	\oplus U(1)
\end{equation}
Such a chain of symmetry breaking could be considered as reflecting the 
evolution of the genetic code, the six amino-acids relative to the codons 
in the irreducible representations obtained after the first breaking (in 
which 64 = 16 + 4 + 20 + 10 + 12 + 2) appearing as primordial amino-acids 
in their approach.  However, the authors were obliged, in order to 
reproduce the actual multiplet pattern, to assume in the final breaking, a 
partial breaking or a ``freezing'' in the sense that the breaking of the 
last $su(2)$ into $U(1)$ does not occur for all the multiplets.  As an 
example, such a freezing has to be imposed to the sextets corresponding to 
Leu and Ser, which otherwise would decompose into three doublets.  In the 
same way, freezing will forbid the doublets related to Lys and Cys to split 
into singlets.

In a second further paper, dated 1997 \cite{FHH}, a refinement of this 
approach has been considered, with the use of Lie groups instead of Lie 
algebras: then, global properties, for example non connexity of $O(2) = 
U(1) \times \ZZ_2$, can be exploited.  In this context, the authors 
proposed another chain of breaking starting with the exceptional group 
$G_2$, which also allows a 64 dimensional irreducible representation.  But 
here again, the freezing pathology cannot be avoided.

One can also mention the work of \cite{KSW} where the unifying algebra 
before breaking is $so(14)$.
  
Meantime (1997), interpreting the double origin of the nucleotides, each 
arising either from purine or from pyrimidine, as a $\ZZ_{2}$-grading a 
supersymmetric model was proposed \cite{BTJ}, involving superalgebras for 
such a program.  The $\ZZ_2$-grading specific of a simple superalgebra is 
there used to separate purine and pyrimidine: indeed, by putting the four 
nucleotids in the the 4 dimensional representation of $su(2/1)$ one can 
confer to the A and G purines (R) an even grading, and to the C and U 
pyrimidines (Y) an odd grading; note that the R states are then in the 
$su(2)$ doublet and the Y ones $su(2)$ singlets.  The notion of polarity 
spin is also introduced, allowing to distinguish the C and G nucleotides 
with two locally polarized sites, from the A and U ones with three 
polarized sites: the C and G (resp.  A and U) will be assigned in a doublet 
(resp.  in singlets) of another $su(2)$.  Then the authors consider the sum 
of algebras: $su(2) \oplus su(2) \oplus su(2|1)$ with the first (second) 
$su(2)$ acting as polarity spin on the first (second) nucleotid of a codon, 
and the $su(2|1)$ acting on the third nucleotid only.  Moreover the two 
$su(2)$ would act in an alternating way on the first and second position, 
that is as $1/2$, $-1/2$ and $-1/2$, $1/2$.  This sum of algebras can be 
embedded in the superalgebra $su(6|1)$, which admits a 64 dimensional 
irreducible representation, and could be also used for a superalgebraic 
approach to the genetic code evolution, with the chain of symmetry 
breaking:
\begin{eqnarray}
	\lefteqn{su(6|1) \supset su(2) \oplus su(3|1) \supset su(2) \oplus 
	su(2) \oplus su(2|1) \supset U(1) \oplus U (1) \oplus su(2|1)} 
	\hspace*{100mm} \nonumber \\
	&& \supset U(1) \oplus U(1) \oplus gl(1|1)
\end{eqnarray}
Again the problem of freezing, that is the last breaking applies to some 
but not all the multiplets, is present with this choice of (super)algebras.

It seems necessary to remark that in this proposal which implies 
(super)algebras acting in the same time on nucleotides and on codons -- one 
must say in a rather complicated way -- the nucleotides cannot appear as 
building blocks from which one algebraically constructs the codons, by 
performing tensorial products of representations, as is the case of our 
model.  In fact, the problem of ordering the nucleotides inside a codon 
forbids this natural way of proceeding as long as only usual 
(super)algebras are involved.  Note that it is the limit of quantum 
algebras that we use in our approach: then, we have at hand the so-called 
crystal bases, which exactly solve the ordering problem.

In a last month preprint, two authors of the same team \cite{BJ} proposed 
to fit biophysical properties of nucleic acids by constructing polynomials 
in 6 coordinates in the 64 dimensional codon space.  As already mentioned 
in sect.  \ref{fits}, the two coordinates they associate to each nucleotide 
is direcetly related to the nucleotide eigenvalues of our model.  The 
authors present their computations as independent of a particular choice of 
algebra or superalgebra as long as the underlying algebra is of rank 6 -- 
which is in particular the dimension of the Cartan subalgebra of $su(6|1)$ 
-- and admits a 64 dimensional irreducible representation.  We note that 
our model does allow to calculate the biophysical quantities considered in 
ref.  \cite{BJ} without the constraint on representations, but more 
importantly, with only a two rank algebra.

A detailed and systematic study of superalgebras and superalgebra breaking 
chains has been performed by the authors of \cite{FS}: it is the 
orthosymplectic $osp(5|2)$ superalgebra which emerges from their algebraic 
analysis.

Finally, it is amazing to remark that, just a few years after the the 
concept of genetic code was formulated, an attempt to give a mathematical 
description of its properties was started by the russian physicist Yu.  B. 
Rumer \cite{R}.  Indeed he remarked that the 16 \emph{roots}, i.e. the 
combinations of the first two codons, divide in a \emph{strong octet} which 
form quartets ou sub-part of sextets and a \emph{weak octet} which form 
doublets, triplets and singlets, attempting to give a systematic 
description of the genetic code.  A few years after, with B.G. 
Konopel'chenko \cite{KR} they formulated the strong assumption that with 
respect to any property of the codons the 16 roots can be gathered into two 
octets with opposite ``charge'', whose positive (negative) value 
respectively characterizes the strong and weak roots.  This description 
comes out naturally in our model, such a charge $Q$ being defined in Eq.  
(5) of sect.  \ref{model}.

\section{Conclusion}
\label{conclusion}

Our model is based on the algebra ${\cal U}_{q \to 0}(sl(2) \oplus sl(2))$ 
that we have chosen for two main characteristics.  First it encodes the 
stereochemical property of a base, and also reflects the complementarity 
rule, by conferring quantum numbers to each nucleotide.  Secondly, it 
admits representation spaces or crystal bases in which an ordered sequence 
of nucleotides or codon can be suitably characterized.  Let us emphasize 
that ${\cal U}_{q \to 0}(sl(2) \oplus sl(2))$ is really neither a Lie 
algebra nor an enveloping deformed algebra.  We still use in a loose sense 
the word algebra, just to emphasize the fact that we use largely the 
mathematical tools of representation space, tensor operators etc.  which 
are typical of the algebraic structures.  Let us add that it is a 
remarkable property of a quantum algebra in the limit $q \to 0$ to admit 
representations, obtained from the tensorial product of basic ones, in 
which each state appears as a unique sequence of ordered basic elements.

In this framework, the correspondence codon/amino-acid is realized by the 
operator ${\cal R}_{c}$, constructed out of the symmetry algebra, and 
acting on codons: the eigenvalues provided by ${\cal R}_{c}$ on two codons 
will be equal or different depending on whether the two codons are 
associated to the same or to two different amino-acids.  It is remarkable 
that this correspondence can be obtained for all the genetic codes and that 
the reading operators have a bulk common to the various genetic codes (the 
prototype reading operator) and differ only for a few additive terms, 
analogous to perturbative terms present in most Hamiltonians describing 
complex physical systems.  Moreover they depend on parameters, presently 
assumed as constants, which in principle can be considered as functions of 
suitable variables.  These feature may be of some interest in the study of 
the evolution of the genetic code, problem which has not yet been tackled 
in our model.

Then, restricting to the case of states made of two nucleotides, the 
experimental values of the free energy, released by base pairing in the 
formation of double stranded nucleic acids, of the hydrophibicity and of 
the hydrophilicity have been fitted with expressions depending respectively 
on 2, 4 and 4 parameters and constructed out of the generators of ${\cal 
U}_{q \to 0}(sl(2) \oplus sl(2))$.
 
The model does not necessarily assign the codons in a multiplet (in 
particular the quartets, sextets and triplet) to the same irreducible 
representation.  Let us remark that the assignments of the codons to the 
different irreducible representations is a straightforward consequence of 
the tensor product, once assigned the nucleotides to the fundamental 
irreducible representation.  This feature is relevant, since it can explain 
the correlation between the branching ratios of the codon usage of 
different codons coding the same amino-acid as discussed in \cite{FSS2} and 
\cite{CFSS}.  Here we have shown that the universal pattern (inside the 
class of vertebrates) of $B_{AG}/B_{UC}$ can simply be reproduced in our 
model.

Moreover our mathematical description of the genetic code allows a 
modelisation of some biological process.  A first step in this direction 
has been presented in sect.  \ref{mutation}, where we have shown that the 
observed data related to the a pyrimidine deletion can be simulated by 
introducing the concept of $q \to 0$ -- or crystal -- tensor operator.  
Finally let us mention some directions for future development of our model.  
Going further in the analysis of the branching ratios, we want to refine 
our analysis and make a more detailed study taking into account the 
dependence on the family of biological species.  Indeed preliminary 
analysis on plants, invertebrates and bacteriae shows that, even if the 
pattern of the correlation is still approximatively present, large 
deviations appear which presumably exhibit evidence that the dependence on 
subclass or family of biological species cannot any more be neglected, 
differently to the case of vertebrates.  A further investigation of the 
possibility of mathematically modelising or simulating biological 
processes, in particular mutations, by crystal tensor operators, is in 
progress.  Other questions are still to be investigated: in particular how 
could the genetic code evolution be reproduced in our model ?

\clearpage 

\begin{center}
\begin{picture}(400,-600)
\multiput(0,0)(75,0){4}{\circle*{5}}
\multiput(0,0)(75,0){4}{\line(3,-2){60}}
\multiput(0,0)(75,0){3}{\line(1,0){75}}
\multiput(60,-40)(75,0){4}{\circle*{5}}
\multiput(60,-40)(75,0){4}{\line(3,-2){60}}
\multiput(60,-40)(75,0){3}{\line(1,0){75}}
\multiput(120,-80)(75,0){4}{\circle*{5}}
\multiput(120,-80)(75,0){4}{\line(3,-2){60}}
\multiput(120,-80)(75,0){3}{\line(1,0){75}}
\multiput(180,-120)(75,0){4}{\circle*{5}}
\multiput(180,-120)(75,0){3}{\line(1,0){75}}
\multiput(150,0)(60,-40){4}{\thicklines\line(1,0){75}}
\multiput(150,1)(60,-40){4}{\thicklines\line(1,0){75}}
\put(0,10){\makebox(0.4,0.6){CCC}}
\put(75,10){\makebox(0.4,0.6){UCC}}
\put(150,10){\makebox(0.4,0.6){UUC}}
\put(225,10){\makebox(0.4,0.6){UUU}}
\put(60,-30){\makebox(0.4,0.6){GCC}}
\put(135,-30){\makebox(0.4,0.6){ACC}}
\put(210,-30){\makebox(0.4,0.6){AUC}}
\put(285,-30){\makebox(0.4,0.6){AUU}}
\put(120,-70){\makebox(0.4,0.6){GGC}}
\put(195,-70){\makebox(0.4,0.6){AGC}}
\put(270,-70){\makebox(0.4,0.6){AAC}}
\put(345,-70){\makebox(0.4,0.6){AAU}}
\put(180,-110){\makebox(0.4,0.6){GGG}}
\put(255,-110){\makebox(0.4,0.6){AGG}}
\put(330,-110){\makebox(0.4,0.6){AAG}}
\put(405,-110){\makebox(0.4,0.6){AAA}}
\multiput(0,-180)(60,-40){4}{\circle*{5}}
\multiput(0,-180)(60,-40){3}{\line(3,-2){60}}
\multiput(75,-180)(60,-40){4}{\circle*{5}}
\multiput(75,-180)(60,-40){3}{\line(3,-2){60}}
\multiput(0,-180)(60,-40){4}{\line(1,0){75}}
\put(0,-170){\makebox(0.4,0.6){CCU}}
\put(75,-170){\makebox(0.4,0.6){UCU}}
\put(60,-210){\makebox(0.4,0.6){GCU}}
\put(135,-210){\makebox(0.4,0.6){ACU}}
\put(120,-250){\makebox(0.4,0.6){GGU}}
\put(195,-250){\makebox(0.4,0.6){AGU}}
\put(180,-290){\makebox(0.4,0.6){GGA}}
\put(255,-290){\makebox(0.4,0.6){AGA}}
\put(120,-80){\thicklines\line(3,-2){60}}
\put(120,-81){\thicklines\line(3,-2){60}}
\put(120,-260){\thicklines\line(3,-2){60}}
\put(120,-261){\thicklines\line(3,-2){60}}
\multiput(0,-360)(75,0){4}{\circle*{5}}
\multiput(0,-360)(75,0){4}{\line(3,-2){60}}
\multiput(0,-360)(75,0){3}{\line(1,0){75}}
\multiput(60,-400)(75,0){4}{\circle*{5}}
\multiput(60,-400)(75,0){3}{\line(1,0){75}}
\put(150,-360){\thicklines\line(1,0){75}}
\put(150,-361){\thicklines\line(1,0){75}}
\put(0,-350){\makebox(0.4,0.6){CCG}}
\put(75,-350){\makebox(0.4,0.6){UCG}}
\put(150,-350){\makebox(0.4,0.6){UUG}}
\put(225,-350){\makebox(0.4,0.6){UUA}}
\put(60,-390){\makebox(0.4,0.6){GCG}}
\put(135,-390){\makebox(0.4,0.6){ACG}}
\put(210,-390){\makebox(0.4,0.6){AUG}}
\put(285,-390){\makebox(0.4,0.6){AUA}}
\multiput(0,-540)(75,0){2}{\circle*{5}}
\multiput(0,-540)(75,0){2}{\line(3,-2){60}}
\multiput(60,-580)(75,0){2}{\circle*{5}}
\multiput(0,-540)(60,-40){2}{\line(1,0){75}}
\put(0,-530){\makebox(0.4,0.6){CCA}}
\put(75,-530){\makebox(0.4,0.6){UCA}}
\put(60,-570){\makebox(0.4,0.6){GCA}}
\put(135,-570){\makebox(0.4,0.6){ACA}}
\multiput(0,0)(75,0){2}{\thicklines\line(0,-1){540}}
\multiput(1,0)(75,0){2}{\thicklines\line(0,-1){540}}
\multiput(60,-40)(75,0){2}{\thicklines\line(0,-1){540}}
\multiput(61,-40)(75,0){2}{\thicklines\line(0,-1){540}}
\multiput(120,-80)(75,0){2}{\thicklines\line(0,-1){180}}
\multiput(121,-80)(75,0){2}{\thicklines\line(0,-1){180}}
\multiput(180,-120)(75,0){2}{\thicklines\line(0,-1){180}}
\multiput(181,-120)(75,0){2}{\thicklines\line(0,-1){180}}
\put(165,-210){\makebox(0.4,0.6){\textit{Gly}}}
\put(210,-170){\makebox(0.4,0.6){\textit{Ser$_{2}$}}}
\put(270,-210){\makebox(0.4,0.6){\textit{Arg$_{2}$}}}
\put(187,10){\makebox(0.4,0.6){\textit{Phe}}}
\put(247,-30){\makebox(0.4,0.6){\textit{Ile}}}
\put(307,-70){\makebox(0.4,0.6){\textit{Asn}}}
\put(367,-110){\makebox(0.4,0.6){\textit{Lys}}}
\put(-15,-270){\makebox(0.4,0.6){\textit{Pro}}}
\put(45,-310){\makebox(0.4,0.6){\textit{Ala}}}
\put(90,-270){\makebox(0.4,0.6){\textit{Ser$_{4}$}}}
\put(150,-310){\makebox(0.4,0.6){\textit{Thr}}}
\put(187,-350){\makebox(0.4,0.6){\textit{Leu$_{2}$}}}
\put(400,-60){\makebox(0.4,0.6){$(3/2,3/2)$}}
\put(400,-240){\makebox(0.4,0.6){$(1/2,3/2)^1$}}
\put(400,-380){\makebox(0.4,0.6){$(3/2,1/2)^1$}}
\put(400,-560){\makebox(0.4,0.6){$(1/2,1/2)^1$}}
\end{picture}
\end{center}

\clearpage

\centerline{Figure 1 (continued)}

\begin{center}
\begin{picture}(400,-640)
\multiput(0,0)(60,-40){4}{\circle*{5}}
\multiput(0,0)(60,-40){3}{\line(3,-2){60}}
\multiput(75,0)(60,-40){4}{\circle*{5}}
\multiput(75,0)(60,-40){3}{\line(3,-2){60}}
\multiput(0,0)(60,-40){4}{\thicklines\line(1,0){75}}
\multiput(0,1)(60,-40){4}{\thicklines\line(1,0){75}}
\put(0,10){\makebox(0.4,0.6){CUC}}
\put(75,10){\makebox(0.4,0.6){CUU}}
\put(60,-30){\makebox(0.4,0.6){GUC}}
\put(135,-30){\makebox(0.4,0.6){GUU}}
\put(120,-70){\makebox(0.4,0.6){GAC}}
\put(195,-70){\makebox(0.4,0.6){GAU}}
\put(180,-110){\makebox(0.4,0.6){GAG}}
\put(255,-110){\makebox(0.4,0.6){GAA}}
\multiput(0,-180)(75,0){2}{\circle*{5}}
\multiput(0,-180)(75,0){2}{\line(3,-2){60}}
\multiput(60,-220)(75,0){2}{\circle*{5}}
\multiput(0,-180)(60,-40){2}{\thicklines\line(1,0){75}}
\multiput(0,-181)(60,-40){2}{\thicklines\line(1,0){75}}
\put(0,-170){\makebox(0.4,0.6){CUG}}
\put(75,-170){\makebox(0.4,0.6){CUA}}
\put(60,-210){\makebox(0.4,0.6){GUG}}
\put(135,-210){\makebox(0.4,0.6){GUA}}
\multiput(0,0)(75,0){2}{\thicklines\line(0,-1){180}}
\multiput(1,0)(75,0){2}{\thicklines\line(0,-1){180}}
\multiput(60,-40)(75,0){2}{\thicklines\line(0,-1){180}}
\multiput(61,-40)(75,0){2}{\thicklines\line(0,-1){180}}
\put(-15,-90){\makebox(0.4,0.6){\textit{Leu$_{4}$}}}
\put(45,-130){\makebox(0.4,0.6){\textit{Val}}}
\put(157,-70){\makebox(0.4,0.6){\textit{Asp}}}
\put(217,-110){\makebox(0.4,0.6){\textit{Glu}}}
\multiput(0,-300)(75,0){4}{\circle*{5}}
\multiput(0,-300)(75,0){4}{\line(3,-2){60}}
\multiput(0,-300)(75,0){3}{\line(1,0){75}}
\multiput(60,-340)(75,0){4}{\circle*{5}}
\multiput(60,-340)(75,0){3}{\line(1,0){75}}
\put(0,-290){\makebox(0.4,0.6){CGC}}
\put(75,-290){\makebox(0.4,0.6){UGC}}
\put(150,-290){\makebox(0.4,0.6){UAC}}
\put(225,-290){\makebox(0.4,0.6){UAU}}
\put(60,-330){\makebox(0.4,0.6){CGG}}
\put(135,-330){\makebox(0.4,0.6){UGG}}
\put(210,-330){\makebox(0.4,0.6){UAG}}
\put(285,-330){\makebox(0.4,0.6){UAA}}
\multiput(0,-300)(0,-180){2}{\thicklines\line(3,-2){60}}
\multiput(0,-301)(0,-180){2}{\thicklines\line(3,-2){60}}
\multiput(150,-300)(60,-40){2}{\thicklines\line(1,0){75}}
\multiput(150,-301)(60,-40){2}{\thicklines\line(1,0){75}}
\multiput(0,-480)(75,0){2}{\circle*{5}}
\put(75,-480){\line(3,-2){60}}
\multiput(60,-520)(75,0){2}{\circle*{5}}
\multiput(0,-480)(60,-40){2}{\line(1,0){75}}
\put(0,-470){\makebox(0.4,0.6){CGU}}
\put(75,-470){\makebox(0.4,0.6){UGU}}
\put(60,-510){\makebox(0.4,0.6){CGA}}
\put(135,-510){\makebox(0.4,0.6){UGA}}
\multiput(0,-300)(75,0){2}{\thicklines\line(0,-1){180}}
\multiput(1,-300)(75,0){2}{\thicklines\line(0,-1){180}}
\put(60,-340){\thicklines\line(0,-1){180}}
\put(61,-340){\thicklines\line(0,-1){180}}
\put(-15,-390){\makebox(0.4,0.6){\textit{Arg$_{4}$}}}
\put(90,-390){\makebox(0.4,0.6){\textit{Cys}}}
\put(187,-290){\makebox(0.4,0.6){\textit{Tyr}}}
\put(247,-330){\makebox(0.4,0.6){\textit{Ter}}}
\put(135,-350){\makebox(0.4,0.6){\textit{Trp}}}
\multiput(0,-600)(75,0){2}{\circle*{5}}
\multiput(0,-600)(75,0){2}{\line(3,-2){60}}
\multiput(60,-640)(75,0){2}{\circle*{5}}
\multiput(0,-600)(60,-40){2}{\thicklines\line(1,0){75}}
\multiput(0,-601)(60,-40){2}{\thicklines\line(1,0){75}}
\put(0,-590){\makebox(0.4,0.6){CAC}}
\put(75,-590){\makebox(0.4,0.6){CAU}}
\put(60,-630){\makebox(0.4,0.6){CAG}}
\put(135,-630){\makebox(0.4,0.6){CAA}}
\put(37,-590){\makebox(0.4,0.6){\textit{His}}}
\put(97,-630){\makebox(0.4,0.6){\textit{Gln}}}
\put(400,-60){\makebox(0.4,0.6){$(1/2,3/2)^2$}}
\put(400,-200){\makebox(0.4,0.6){$(1/2,1/2)^3$}}
\put(400,-320){\makebox(0.4,0.6){$(3/2,1/2)^2$}}
\put(400,-500){\makebox(0.4,0.6){$(1/2,1/2)^2$}}
\put(400,-620){\makebox(0.4,0.6){$(1/2,1/2)^4$}}
\end{picture}
\end{center}

\clearpage

\begin{table}[t]
\caption{The eukariotic code. The upper label denotes different irreducible 
representations.}
\label{tablerep}
\footnotesize
\begin{center}
\begin{tabular}{|cc|cc|cc|cc|}
\hline
codon & a.a. & $J_{H}$ & $J_{V}$ & codon & a.a. & $J_{H}$ & $J_{V}$ \\
\hline
CCC & Pro & 3/2 & 3/2 & UCC & Ser & 3/2 & 3/2 \\
CCU & Pro & (1/2 & 3/2$)^1$ & UCU & Ser & (1/2 & 3/2$)^1$ \\
CCG & Pro & (3/2 & 1/2$)^1$ & UCG & Ser & (3/2 & 1/2$)^1$ \\
CCA & Pro & (1/2 & 1/2$)^1$ & UCA & Ser & (1/2 & 1/2$)^1$ \\
\hline
CUC & Leu & (1/2 & 3/2$)^2$ & UUC & Phe & 3/2 & 3/2 \\
CUU & Leu & (1/2 & 3/2$)^2$ & UUU & Phe & 3/2 & 3/2 \\
CUG & Leu & (1/2 & 1/2$)^3$ & UUG & Leu & (3/2 & 1/2$)^1$ \\
CUA & Leu & (1/2 & 1/2$)^3$ & UUA & Leu & (3/2 & 1/2$)^1$ \\ 
\hline
CGC & Arg & (3/2 & 1/2$)^2$ & UGC & Cys & (3/2 & 1/2$)^2$ \\
CGU & Arg & (1/2 & 1/2$)^2$ & UGU & Cys & (1/2 & 1/2$)^2$ \\
CGG & Arg & (3/2 & 1/2$)^2$ & UGG & Trp & (3/2 & 1/2$)^2$ \\
CGA & Arg & (1/2 & 1/2$)^2$ & UGA & Ter & (1/2 & 1/2$)^2$ \\
\hline
CAC & His & (1/2 & 1/2$)^4$ & UAC & Tyr & (3/2 & 1/2$)^2$ \\
CAU & His & (1/2 & 1/2$)^4$ & UAU & Tyr & (3/2 & 1/2$)^2$ \\
CAG & Gln & (1/2 & 1/2$)^4$ & UAG & Ter & (3/2 & 1/2$)^2$ \\
CAA & Gln & (1/2 & 1/2$)^4$ & UAA & Ter & (3/2 & 1/2$)^2$ \\
\hline
GCC & Ala & 3/2 & 3/2 & ACC & Thr & 3/2 & 3/2 \\
GCU & Ala & (1/2 & 3/2$)^1$ & ACU & Thr & (1/2 & 3/2$)^1$ \\
GCG & Ala & (3/2 & 1/2$)^1$ & ACG & Thr & (3/2 & 1/2$)^1$ \\
GCA & Ala & (1/2 & 1/2$)^1$ & ACA & Thr & (1/2 & 1/2$)^1$ \\
\hline
GUC & Val & (1/2 & 3/2$)^2$ & AUC & Ile & 3/2 & 3/2 \\
GUU & Val & (1/2 & 3/2$)^2$ & AUU & Ile & 3/2 & 3/2 \\
GUG & Val & (1/2 & 1/2$)^3$ & AUG & Met & (3/2 & 1/2$)^1$ \\
GUA & Val & (1/2 & 1/2$)^3$ & AUA & Ile & (3/2 & 1/2$)^1$ \\
\hline
GGC & Gly & 3/2 & 3/2 & AGC & Ser & 3/2 & 3/2 \\
GGU & Gly & (1/2 & 3/2$)^1$ & AGU & Ser & (1/2 & 3/2$)^1$ \\
GGG & Gly & 3/2 & 3/2 & AGG & Arg & 3/2 & 3/2 \\
GGA & Gly & (1/2 & 3/2$)^1$ & AGA & Arg & (1/2 & 3/2$)^1$ \\
\hline
GAC & Asp & (1/2 & 3/2$)^2$ & AAC & Asn & 3/2 & 3/2 \\
GAU & Asp & (1/2 & 3/2$)^2$ & AAU & Asn & 3/2 & 3/2 \\
GAG & Glu & (1/2 & 3/2$)^2$ & AAG & Lys & 3/2 & 3/2 \\
GAA & Glu & (1/2 & 3/2$)^2$ & AAA & Lys & 3/2 & 3/2 \\
\hline
\end{tabular}
\end{center}
\end{table} 

\clearpage

\begin{table}[htbp]
\centering
\caption{Biological species sample used in analysis of sect.  \ref{correl}}
\label{tutu}
\medskip
\begin{tabular}{|rlrr|}
	\hline
	& Species & Number & Number \\
	&  & of sequences & of codons \\
	\hline
	1 & Homo sapiens & 17625 & 8707603 \\
	2 & Rattus norvegicus & 4907 & 2469130 \\
	3 & Gallus gallus & 1592 & 763008 \\
	4 & Xenopus laevis & 1433 & 646214 \\
	5 & Bos taurus & 1382 & 614602 \\
	6 & Oryctolagus cuniculus & 713 & 358447 \\
	7 & Sus scrofa & 658 & 275045 \\
	8 & Danio rerio & 500 & 213258 \\
	9 & Rattus rattus & 342 & 153049 \\
	10 & Canis familiaris & 317 & 142944 \\
	11 & Rattus sp. & 299 & 112039 \\
	12 & Ovis aries & 327 & 101591 \\
	13 & Fugu rubripes & 157 & 95979 \\
	\hline
\end{tabular}
\end{table}

\begin{table}[htbp]
\centering
\caption{$B_{AG}$ ratios for the quartets}
\label{ratioAG}
\medskip
\begin{tabular}{|r|rrrrrrrr|}
	\hline
	& Pro & Ala & Thr & Ser & Val & Leu & Arg & Gly \\
	\hline
	1 & 2.34 & 2.03 & 2.29 & 2.51 & 0.23 & 0.17 & 0.53 & 0.99 \\
	2 & 2.40 & 2.17 & 2.33 & 2.35 & 0.22 & 0.17 & 0.61 & 1.03 \\
	3 & 1.77 & 1.90 & 1.96 & 1.93 & 0.25 & 0.14 & 0.52 & 1.02 \\
	4 & 4.10 & 4.23 & 4.08 & 3.45 & 0.48 & 0.32 & 1.00 & 1.67 \\
	5 & 2.02 & 1.80 & 1.94 & 2.32 & 0.21 & 0.14 & 0.56 & 1.01 \\
	6 & 1.45 & 1.45 & 1.30 & 1.45 & 0.15 & 0.10 & 0.44 & 0.88 \\
	7 & 1.60 & 1.60 & 1.52 & 1.69 & 0.16 & 0.12 & 0.46 & 0.89 \\
	8 & 1.39 & 1.47 & 1.71 & 1.68 & 0.22 & 0.18 & 0.89 & 1.94 \\
	9 & 2.28 & 1.97 & 2.19 & 2.26 & 0.21 & 0.17 & 0.66 & 1.03 \\
	10 & 2.09 & 1.72 & 1.81 & 1.90 & 0.21 & 0.15 & 0.49 & 1.01 \\
	11 & 2.22 & 2.15 & 2.27 & 2.24 & 0.21 & 0.16 & 0.62 & 1.07 \\
	12 & 2.15 & 1.60 & 1.76 & 1.99 & 0.15 & 0.13 & 0.60 & 1.08 \\
	13 & 1.60 & 1.40 & 1.28 & 1.42 & 0.17 & 0.12 & 0.73 & 1.23 \\
	\hline
\end{tabular}
\end{table}

\begin{table}[htbp]
\centering
\caption{$B_{UC}$ ratios for the quartets}
\label{ratioUC}
\medskip
\begin{tabular}{|r|rrrrrrrr|}
	\hline
	& Pro & Ala & Thr & Ser & Val & Leu & Arg & Gly \\
	\hline
	1 & 0.85 & 0.64 & 0.64 & 0.82 & 0.72 & 0.64 & 0.43 & 0.47 \\
	2 & 0.91 & 0.69 & 0.61 & 0.78 & 0.59 & 0.57 & 0.48 & 0.49 \\
	3 & 0.75 & 0.80 & 0.69 & 0.77 & 0.84 & 0.64 & 0.45 & 0.51 \\
	4 & 1.27 & 1.15 & 1.05 & 1.17 & 1.26 & 1.24 & 0.98 & 0.87 \\
	5 & 0.78 & 0.61 & 0.57 & 0.79 & 0.65 & 0.57 & 0.41 & 0.47 \\
	6 & 0.62 & 0.47 & 0.46 & 0.54 & 0.51 & 0.43 & 0.29 & 0.34 \\
	7 & 0.68 & 0.54 & 0.49 & 0.65 & 0.50 & 0.47 & 0.33 & 0.38 \\
	8 & 1.02 & 0.88 & 0.69 & 0.83 & 0.82 & 0.64 & 0.60 & 0.68 \\
	9 & 0.88 & 0.71 & 0.59 & 0.76 & 0.59 & 0.57 & 0.50 & 0.51 \\
	10 & 0.76 & 0.61 & 0.57 & 0.76 & 0.56 & 0.55 & 0.37 & 0.53 \\
	11 & 0.94 & 0.69 & 0.58 & 0.83 & 0.55 & 0.55 & 0.47 & 0.49 \\
	12 & 0.70 & 0.53 & 0.45 & 0.73 & 0.50 & 0.46 & 0.41 & 0.43 \\
	13 & 0.77 & 0.68 & 0.55 & 0.71 & 0.60 & 0.49 & 0.57 & 0.64 \\
	\hline
\end{tabular}
\end{table}

\begin{table}[htbp]
\centering
\caption{$B_{AG}/B_{UC}$ ratios for the quartets}
\label{ratio}
\medskip
\begin{tabular}{|r|rrrrrrrr|}
	\hline
	& Pro & Ala & Thr & Ser & Val & Leu & Arg & Gly \\
	\hline
	1 & 2.75 & 3.15 & 3.57 & 3.05 & 0.32 & 0.26 & 1.25 & 2.11 \\
	2 & 2.63 & 3.15 & 3.81 & 3.02 & 0.38 & 0.30 & 1.28 & 2.10 \\
	3 & 2.38 & 2.38 & 2.83 & 2.50 & 0.30 & 0.21 & 1.14 & 2.00 \\
	4 & 3.22 & 3.69 & 3.89 & 2.96 & 0.38 & 0.25 & 1.02 & 1.92 \\
	5 & 2.60 & 2.96 & 3.40 & 2.92 & 0.32 & 0.25 & 1.36 & 2.17 \\
	6 & 2.33 & 3.08 & 2.80 & 2.67 & 0.29 & 0.24 & 1.55 & 2.60 \\
	7 & 2.34 & 2.97 & 3.11 & 2.60 & 0.32 & 0.25 & 1.38 & 2.36 \\
	8 & 1.36 & 1.68 & 2.48 & 2.03 & 0.27 & 0.27 & 1.48 & 2.87 \\
	9 & 2.58 & 2.78 & 3.68 & 2.98 & 0.36 & 0.31 & 1.32 & 2.00 \\
	10 & 2.74 & 2.82 & 3.17 & 2.51 & 0.38 & 0.28 & 1.32 & 1.91 \\
	11 & 2.36 & 3.14 & 3.93 & 2.71 & 0.38 & 0.29 & 1.34 & 2.18 \\
	12 & 3.08 & 3.03 & 3.92 & 2.72 & 0.30 & 0.28 & 1.45 & 2.52 \\
	13 & 2.09 & 2.06 & 2.31 & 2.01 & 0.27 & 0.24 & 1.28 & 1.92 \\
	\hline
\end{tabular}
\end{table}

\begin{table}
\centering
\caption{$F$ functions appearing in the $B_{AG}/B_{UC}$ ratios}
\label{irreps}
\begin{displaymath}
\begin{array}{|cccc|}
	\hline
	\mathrm{Pro} & \mathrm{Ala} & \mathrm{Thr} & \mathrm{Ser} \\
	\hline
	\displaystyle \frac{F_{AG}\Big((\half,\half)^1;(\thalf,\half)^1\Big)} 
	{F_{UC}\Big((\half,\thalf)^1;(\thalf,\thalf)\Big)} & \displaystyle 
	\frac{F_{AG}\Big((\half,\half)^1;(\thalf,\half)^1\Big)} 
	{F_{UC}\Big((\half,\thalf)^1;(\thalf,\thalf)\Big)} & \displaystyle 
	\frac{F_{AG}\Big((\half,\half)^1;(\thalf,\half)^1\Big)} 
	{F_{UC}\Big((\half,\thalf)^1;(\thalf,\thalf)\Big)} & \displaystyle 
	\frac{F_{AG}\Big((\half,\half)^1;(\thalf,\half)^1\Big)} 
	{F_{UC}\Big((\half,\thalf)^1;(\thalf,\thalf)\Big)} \\
	\hline
	\hline
	\mathrm{Val} & \mathrm{Leu} & \mathrm{Arg} & \mathrm{Gly} \\
	\hline
	\displaystyle \frac{F_{AG}\Big((\half,\half)^3;(\half,\half)^3\Big)} 
	{F_{UC}\Big((\half,\thalf)^2;(\half,\thalf)^2\Big)} & 
	\displaystyle \frac{F_{AG}\Big((\half,\half)^3;(\half,\half)^3\Big)} 
	{F_{UC}\Big((\half,\thalf)^2;(\half,\thalf)^2\Big)} & 
	\displaystyle \frac{F_{AG}((\half,\half)^2;(\thalf,\half)^2\Big)} 
	{F_{UC}\Big((\half,\half)^2;(\thalf,\half)^2\Big)} & 
	\displaystyle \frac{F_{AG}\Big((\half,\thalf)^1;(\thalf,\thalf)\Big)} 
	{F_{UC}\Big((\half,\thalf)^2;(\thalf,\thalf)\Big)} \\
	\hline
\end{array}
\end{displaymath}
\end{table}

\clearpage

\begin{table}
\centering
\caption{Amino-acid content of the $\otimes^3 (\half,\half)$ 
representations}
\label{euka}
\begin{eqnarray*}
	&(\sfrac{3}{2},\sfrac{3}{2}) \equiv \left( 
	\begin{array}{ccccccc} 
		P-\texttt{Pro} && S-\texttt{Ser} && F-\texttt{Phe} && F-\texttt{Phe} 
		\\
		A-\texttt{Ala} && T-\texttt{Thr} && I-\texttt{Ile} && I-\texttt{Ile} 
		\\
		G-\texttt{Gly} && S-\texttt{Ser} && N-\texttt{Asn} && N-\texttt{Asn} 
		\\
		G-\texttt{Gly} && R-\texttt{Arg} && K-\texttt{Lys} && K-\texttt{Lys} 
		\\
	\end{array} 
	\right)& \\
	&& \\
	&& \\
	&(\sfrac{3}{2},\sfrac{1}{2})^{1} \equiv \left( 
	\begin{array}{ccccccc} 
		P-\texttt{Pro} && S-\texttt{Ser} && L-\texttt{Leu} && L-\texttt{Leu} 
		\\
		A-\texttt{Ala} && T-\texttt{Thr} && M-\texttt{Met} && I-\texttt{Ile} 
		\\
	\end{array} 
	\right)& \\
	&& \\
	&& \\
	&(\sfrac{3}{2},\sfrac{1}{2})^{2} \equiv \left( 
	\begin{array}{ccccccc} 
		R-\texttt{Arg} && C-\texttt{Cys} && Y-\texttt{Tyr} && Y-\texttt{Tyr} 
		\\
		R-\texttt{Arg} && W-\texttt{Trp} && \texttt{Ter} && \texttt{Ter} \\
	\end{array} 
	\right)& \\
	&& \\
	&& \\
	&(\sfrac{1}{2},\sfrac{3}{2})^{1} \equiv \left( 
	\begin{array}{ccc}
		P-\texttt{Pro} && S-\texttt{Ser} \\
		A-\texttt{Ala} && T-\texttt{Thr} \\
		G-\texttt{Gly} && S-\texttt{Ser} \\
		G-\texttt{Gly} && R-\texttt{Arg} \\
	\end{array}
	\right)& \\
	&& \\
	&& \\
	&(\sfrac{1}{2},\sfrac{3}{2})^{2} \equiv \left(
	\begin{array}{ccc}
		L-\texttt{Leu} && L-\texttt{Leu} \\
		V-\texttt{Val} && V-\texttt{Val} \\
		D-\texttt{Asp} && D-\texttt{Asp} \\
		E-\texttt{Glu} && E-\texttt{Glu} \\
	\end{array}
	\right)& \\
	&& \\
	&& \\
	&(\sfrac{1}{2},\sfrac{1}{2})^{1} \equiv \left( 
	\begin{array}{ccc}
		P-\texttt{Pro} && S-\texttt{Ser} \\ 
		A-\texttt{Ala} && T-\texttt{Thr} \\ 
	\end{array} 
	\right)& \\
	&& \\
	&& \\
	&(\sfrac{1}{2},\sfrac{1}{2})^{2} \equiv \left( 
	\begin{array}{ccc}
		R-\texttt{Arg} && C-\texttt{Cys} \\ 
		R-\texttt{Arg} && \texttt{Ter} \\ 
	\end{array} 
	\right)& \\
	&& \\
	&& \\
	&(\sfrac{1}{2},\sfrac{1}{2})^{3} \equiv \left( 
	\begin{array}{ccc}
		L-\texttt{Leu} && L-\texttt{Leu} \\ 
		V-\texttt{Val} && V-\texttt{Val} \\ 
	\end{array} 
	\right)& \\
	&& \\
	&& \\
	&(\sfrac{1}{2},\sfrac{1}{2})^{4} \equiv \left( 
	\begin{array}{ccc}
		H-\texttt{His} && H-\texttt{His} \\ 
		Q-\texttt{Gln} && Q-\texttt{Gln} \\ 
	\end{array} 
	\right)&
\end{eqnarray*}
\end{table}

\clearpage

\begin{table}
\centering
\caption{Four-fold tensor product of the $(\half,\half)$ representation of 
${\cal U}_{q \rightarrow 0}(sl(2) \oplus sl(2))$}
\end{table}

\begin{eqnarray*}
	(\half,\half) \, \otimes \, (\half,\half) \, \otimes \, (\half,\half) 
	\, \otimes \, (\half,\half) &=& (\half,\half) \otimes \Big[ 
	(\thalf,\thalf) \oplus 2 \, (\thalf,\half) \oplus 2 \, (\half,\thalf) 
	\oplus 4 \, (\half,\half) \Big] \\
	&=& (2,2) \, \oplus \, 3 \, (2,1) \, \oplus \, 3 \, (1,2) \, \oplus \, 
	9 \, (1,1) \, \oplus \, 2 \, (2,0) \\
	&& \oplus \, 2 \, (0,2) \, \oplus \, 6 \, (1,0) \, \oplus \, 6 \, (0,1) 
	\, \oplus \, 4 \, (0,0)
\end{eqnarray*}
One has (The upper label denotes different irreducible representations):

\begin{displaymath}
	(\half,\half) \, \otimes \, (\thalf,\thalf) = (2,2) \, \oplus \, 
	(2,1)^1 \, \oplus \, (1,2)^1 \, \oplus \, (1,1)^1
\end{displaymath}
where
\begin{displaymath}
\begin{array}{ccc}
	(2,2) = && (1,2)^1 = \\
	\left( 
	\begin{array}{ccccc} 
		\mbox{CCCC} & \mbox{UCCC} & \mbox{UUCC} & \mbox{UUUC} & \mbox{UUUU} \\
		\mbox{GCCC} & \mbox{ACCC} & \mbox{AUCC} & \mbox{AUUC} & \mbox{AUUU} \\
		\mbox{GGCC} & \mbox{AGCC} & \mbox{AACC} & \mbox{AAUC} & \mbox{AAUU} \\
		\mbox{GGGC} & \mbox{AGGC} & \mbox{AAGC} & \mbox{AAAC} & \mbox{AAAU} \\
		\mbox{GGGG} & \mbox{AGGG} & \mbox{AAGG} & \mbox{AAAG} & \mbox{AAAA} \\
	\end{array}
	\right)
	&&
	\left(
	\begin{array}{ccccc} 
		\mbox{CUCC} & \mbox{CUUC} & \mbox{CUUU} \\
		\mbox{GUCC} & \mbox{GUUC} & \mbox{GUUU} \\
		\mbox{GACC} & \mbox{GAUC} & \mbox{GAUU} \\
		\mbox{GAGC} & \mbox{GAAC} & \mbox{GAAU} \\
		\mbox{GAGG} & \mbox{GAAG} & \mbox{GAAA} \\
	\end{array}
	\right) \\
	&& \\
	(2,1)^1 = && (1,1)^1 = \\
	\left( 
	\begin{array}{ccccc} 
		\mbox{CGCC} & \mbox{UGCC} & \mbox{UACC} & \mbox{UAUC} & \mbox{UAUU} \\
		\mbox{CGGC} & \mbox{UGGC} & \mbox{UAGC} & \mbox{UAAC} & \mbox{UAAU} \\
		\mbox{CGGG} & \mbox{UGGG} & \mbox{UAGG} & \mbox{UAAG} & \mbox{UAAA} \\
	\end{array}
	\right)
	&&
	\left( 
	\begin{array}{ccccc} 
		\mbox{CACC} & \mbox{CAUC} & \mbox{CAUU} \\
		\mbox{CAGC} & \mbox{CAAC} & \mbox{CAAU} \\
		\mbox{CAGG} & \mbox{CAAG} & \mbox{CAAA} \\
	\end{array}
	\right) \\
\end{array}
\end{displaymath}

\hrule

\begin{displaymath}
	(\half,\half) \, \otimes \, (\thalf,\half)^1 = (2,1)^2 \, \oplus \, 
	(2,0)^1 \, \oplus \, (1,1)^2 \, \oplus \, (1,0)^1
\end{displaymath}
where
\begin{displaymath}
\begin{array}{ccc}
	(2,1)^2 = && (1,1)^2 = \\
	\left( 
	\begin{array}{ccccc} 
		\mbox{CCCG} & \mbox{UCCG} & \mbox{UUCG} & \mbox{UUUG} & \mbox{UUUA} \\
		\mbox{GCCG} & \mbox{ACCG} & \mbox{AUCG} & \mbox{AUUG} & \mbox{AUUA} \\
		\mbox{GGCG} & \mbox{AGCG} & \mbox{AACG} & \mbox{AAUG} & \mbox{AAUA} \\
	\end{array}
	\right)
	&&
	\left(
	\begin{array}{ccccc} 
		\mbox{CUCG} & \mbox{CUUG} & \mbox{CUUA} \\
		\mbox{GUCG} & \mbox{GUUG} & \mbox{GUUA} \\
		\mbox{GACG} & \mbox{GAUG} & \mbox{GAUA} \\
	\end{array}
	\right) \\
	&& \\
	(2,0)^1 = && (1,0)^1 = \\
	\left( 
	\begin{array}{ccccc} 
		\mbox{CGCG} & \mbox{UGCG} & \mbox{UACG} & \mbox{UAUG} & \mbox{UAUA} \\
	\end{array}
	\right)
	&&
	\left( 
	\begin{array}{ccccc} 
		\mbox{CACG} & \mbox{CAUG} & \mbox{CAUA} \\
	\end{array}
	\right) \\
\end{array}
\end{displaymath}

\hrule

\begin{displaymath}
	(\half,\half) \, \otimes \, (\thalf,\half)^2 = (2,1)^3 \, \oplus \, 
	(2,0)^2 \, \oplus \, (1,1)^3 \, \oplus \, (1,0)^2
\end{displaymath}
where
\begin{displaymath}
\begin{array}{ccc}
	(2,1)^3 = && (1,1)^3 = \\
	\left( 
	\begin{array}{ccccc} 
		\mbox{CCGC} & \mbox{UCGC} & \mbox{UUGC} & \mbox{UUAC} & \mbox{UUAU} \\
		\mbox{GCGC} & \mbox{ACGC} & \mbox{AUGC} & \mbox{AUAC} & \mbox{AUAU} \\
		\mbox{GCGG} & \mbox{ACGG} & \mbox{AUGG} & \mbox{AUAG} & \mbox{AUAA} \\
	\end{array}
	\right)
	&&
	\left(
	\begin{array}{ccccc} 
		\mbox{CUGC} & \mbox{CUAC} & \mbox{CUAU} \\
		\mbox{GUGC} & \mbox{GUAC} & \mbox{GUAU} \\
		\mbox{GUGG} & \mbox{GUAG} & \mbox{GUAA} \\
	\end{array}
	\right) \\
	&& \\
	(2,0)^2 = && (1,0)^2 = \\
	\left( 
	\begin{array}{ccccc} 
		\mbox{CCGG} & \mbox{UCGG} & \mbox{UUGG} & \mbox{UUAG} & \mbox{UUAA} \\
	\end{array}
	\right)
	&&
	\left( 
	\begin{array}{ccccc} 
		\mbox{CUGG} & \mbox{CUAG} & \mbox{CUAA} \\
	\end{array}
	\right) \\
\end{array}
\end{displaymath}


\begin{displaymath}
	(\half,\half) \, \otimes \, (\half,\thalf)^1 = (1,2)^2 \, \oplus \, 
	(0,2)^1 \, \oplus \, (1,1)^4 \, \oplus \, (0,1)^1
\end{displaymath}
where
\begin{displaymath}
\begin{array}{ccc}
	(1,2)^2 = \left( 
	\begin{array}{ccccc} 
		\mbox{CCCU} & \mbox{UCCU} & \mbox{UUCU} \\
		\mbox{GCCU} & \mbox{ACCU} & \mbox{AUCU} \\
		\mbox{GGCU} & \mbox{AGCU} & \mbox{AACU} \\
		\mbox{GGGU} & \mbox{AGGU} & \mbox{AAGU} \\
		\mbox{GGGA} & \mbox{AGGA} & \mbox{AAGA} \\
	\end{array}
	\right)
	&&
	(0,2)^1 = \left(
	\begin{array}{ccccc} 
		\mbox{CUCU} \\
		\mbox{GUCU} \\
		\mbox{GACU} \\
		\mbox{GAGU} \\
		\mbox{GAGA} \\
	\end{array}
	\right) \\
	&& \\
	(1,1)^4 = \left( 
	\begin{array}{ccccc} 
		\mbox{CGCU} & \mbox{UGCU} & \mbox{UACU} \\
		\mbox{CGGU} & \mbox{UGGU} & \mbox{UAGU} \\
		\mbox{CGGA} & \mbox{UGGA} & \mbox{UAGA} \\
	\end{array}
	\right)
	&&
	(0,1)^1 = \left( 
	\begin{array}{ccccc} 
		\mbox{CACU} \\
		\mbox{CAGU} \\
		\mbox{CAGA} \\
	\end{array}
	\right) \\
\end{array}
\end{displaymath}

\hrule

\begin{displaymath}
	(\half,\half) \, \otimes \, (\half,\thalf)^2 = (1,2)^3 \, \oplus \, 
	(0,2)^2 \, \oplus \, (1,1)^5 \, \oplus \, (0,1)^2
\end{displaymath}
where
\begin{displaymath}
\begin{array}{ccc}
	(1,2)^3 = \left( 
	\begin{array}{ccccc} 
		\mbox{CCUC} & \mbox{UCUC} & \mbox{UCUU} \\
		\mbox{GCUC} & \mbox{ACUC} & \mbox{ACUU} \\
		\mbox{GGUC} & \mbox{AGUC} & \mbox{AGUU} \\
		\mbox{GGAC} & \mbox{AGAC} & \mbox{AGAU} \\
		\mbox{GGAG} & \mbox{AGAG} & \mbox{AGAA} \\
	\end{array}
	\right)
	&&
	(0,2)^2 = \left(
	\begin{array}{ccccc} 
		\mbox{CCUU} \\
		\mbox{GCUU} \\
		\mbox{GGUU} \\
		\mbox{GGAU} \\
		\mbox{GGAA} \\
	\end{array}
	\right) \\
	&& \\
	(1,1)^5 = \left( 
	\begin{array}{ccccc} 
		\mbox{CGUC} & \mbox{UGUC} & \mbox{UGUU} \\
		\mbox{CGAC} & \mbox{UGAC} & \mbox{UGAU} \\
		\mbox{CGAG} & \mbox{UGAG} & \mbox{UGAA} \\
	\end{array}
	\right)
	&&
	(0,1)^2 = \left( 
	\begin{array}{ccccc} 
		\mbox{CGUU} \\
		\mbox{CGAU} \\
		\mbox{CGAA} \\
	\end{array}
	\right) \\
\end{array}
\end{displaymath}

\hrule

\begin{displaymath}
	(\half,\half) \, \otimes \, (\half,\half)^1 = (1,1)^6 \, \oplus \, 
	(1,0)^3 \, \oplus \, (0,1)^3 \, \oplus \, (0,0)^1
\end{displaymath}
where
\begin{displaymath}
\begin{array}{ccc}
	(1,1)^6 = \left( 
	\begin{array}{ccccc} 
		\mbox{CCCA} & \mbox{UCCA} & \mbox{UUCA} \\
		\mbox{GCCA} & \mbox{ACCA} & \mbox{AUCA} \\
		\mbox{GGCA} & \mbox{AGCA} & \mbox{AACA} \\
	\end{array}
	\right)
	&&
	(0,1)^3 = \left(
	\begin{array}{ccccc} 
		\mbox{CUCA} \\
		\mbox{GUCA} \\
		\mbox{GACA} \\
	\end{array}
	\right) \\
	&& \\
	(1,0)^3 = \left( 
	\begin{array}{ccccc} 
		\mbox{CGCA} & \mbox{UGCA} & \mbox{UACA} \\
	\end{array}
	\right)
	&&
	(0,0)^1 = \left( 
	\begin{array}{ccccc} 
		\mbox{CACA} \\
	\end{array}
	\right) \\
\end{array}
\end{displaymath}

\hrule

\begin{displaymath}
	(\half,\half) \, \otimes \, (\half,\half)^2 = (1,1)^7 \, \oplus \, 
	(1,0)^4 \, \oplus \, (0,1)^4 \, \oplus \, (0,0)^2
\end{displaymath}
where
\begin{displaymath}
\begin{array}{ccc}
	(1,1)^7 = \left( 
	\begin{array}{ccccc} 
		\mbox{CCGU} & \mbox{UCGU} & \mbox{UUGU} \\
		\mbox{GCGU} & \mbox{ACGU} & \mbox{AUGU} \\
		\mbox{GCGA} & \mbox{ACGA} & \mbox{AUGA} \\
	\end{array}
	\right)
	&&
	(0,1)^4 = \left(
	\begin{array}{ccccc} 
		\mbox{CUGU} \\
		\mbox{GUGU} \\
		\mbox{GUGA} \\
	\end{array}
	\right) \\
	&& \\
	(1,0)^4 = \left( 
	\begin{array}{ccccc} 
		\mbox{CCGA} & \mbox{UCGA} & \mbox{UUGA} \\
	\end{array}
	\right)
	&&
	(0,0)^2 = \left( 
	\begin{array}{ccccc} 
		\mbox{CUGA} \\
	\end{array}
	\right) \\
\end{array}
\end{displaymath}

\newpage

\begin{displaymath}
	(\half,\half) \, \otimes \, (\half,\half)^3 = (1,1)^8 \, \oplus \, 
	(1,0)^5 \, \oplus \, (0,1)^5 \, \oplus \, (0,0)^3
\end{displaymath}
where
\begin{displaymath}
\begin{array}{ccc}
	(1,1)^8 = \left( 
	\begin{array}{ccccc} 
		\mbox{CCUG} & \mbox{UCUG} & \mbox{UCUA} \\
		\mbox{GCUG} & \mbox{ACUG} & \mbox{ACUA} \\
		\mbox{GGUG} & \mbox{AGUG} & \mbox{AGUA} \\
	\end{array}
	\right)
	&&
	(0,1)^5 = \left(
	\begin{array}{ccccc} 
		\mbox{CCUA} \\
		\mbox{GCUA} \\
		\mbox{GGUA} \\
	\end{array}
	\right) \\
	&& \\
	(1,0)^5 = \left( 
	\begin{array}{ccccc} 
		\mbox{CGUG} & \mbox{UGUG} & \mbox{UGUA} \\
	\end{array}
	\right)
	&&
	(0,0)^3 = \left( 
	\begin{array}{ccccc} 
		\mbox{CGUA} \\
	\end{array}
	\right) \\
\end{array}
\end{displaymath}

\hrule

\begin{displaymath}
	(\half,\half) \, \otimes \, (\half,\half)^4 = (1,1)^9 \, \oplus \, 
	(1,0)^6 \, \oplus \, (0,1)^6 \, \oplus \, (0,0)^4
\end{displaymath}
where
\begin{displaymath}
\begin{array}{ccc}
	(1,1)^9 = \left( 
	\begin{array}{ccccc} 
		\mbox{CCAC} & \mbox{UCAC} & \mbox{UCAU} \\
		\mbox{GCAC} & \mbox{ACAC} & \mbox{ACAU} \\
		\mbox{GCAG} & \mbox{ACAG} & \mbox{ACAA} \\
	\end{array}
	\right)
	&&
	(0,1)^6 = \left(
	\begin{array}{ccccc} 
		\mbox{CCAU} \\
		\mbox{GCAU} \\
		\mbox{GCAA} \\
	\end{array}
	\right) \\
	&& \\
	(1,0)^6 = \left( 
	\begin{array}{ccccc} 
		\mbox{CCAG} & \mbox{UCAG} & \mbox{UCAA} \\
	\end{array}
	\right)
	&&
	(0,0)^4 = \left( 
	\begin{array}{ccccc} 
		\mbox{CCAA} \\
	\end{array}
	\right) \\
\end{array}
\end{displaymath}

\end{document}